\documentclass[amsmath, amssymb, english, aps, prb, showpacs, twocolumn, reprint, floatfix, superscriptaddress,longbibliography]{revtex4-1}
\usepackage[english]{babel}
\usepackage{graphicx}
\usepackage[usenames,dvipsnames]{color}
\usepackage[normalem]{ulem}
\usepackage{hyperref}
\usepackage{float}
\usepackage{color}

\usepackage{tikz}

\newcommand{\bk}{\mathbf{k}}
\newcommand{\bK}{\mathbf{K}}
\newcommand{\bq}{\mathbf{q}}
\newcommand{\bQ}{\mathbf{Q}}
\newcommand{\re}{\mathrm{Re}}
\newcommand{\im}{\mathrm{Im}}

\newcommand{\pdagger}{{\phantom{\dagger}}}

\renewcommand{\approx}{\simeq}

\renewcommand{\vec}[1]{\boldsymbol{#1}}

\newcommand{\bismtwolayer}{Bi$_2$Sr$_2$CaCu$_2$O$_{8+\delta}$\,}

\begin{document}

\author{Wei Wu}
\affiliation{Centre de Physique Th\'eorique, Ecole Polytechnique, CNRS, Universit\'e Paris-Saclay, 91128 Palaiseau, France}
\affiliation{Coll\`ege de France, 11 place Marcelin Berthelot, 75005 Paris, France}
\author{Mathias S.~Scheurer}
\affiliation{Department of Physics, Harvard University, Cambridge MA 02138, USA}
\author{Shubhayu Chatterjee}
\affiliation{Department of Physics, Harvard University, Cambridge MA 02138, USA}
\author{Subir Sachdev}
\affiliation{Department of Physics, Harvard University, Cambridge MA 02138, USA}
\affiliation{Perimeter Institute for Theoretical Physics, Waterloo, Ontario, Canada N2L 2Y5}
\affiliation{Department of Physics, Stanford University, Stanford, CA 94305}
\author{Antoine Georges}
\affiliation{Coll\`ege de France, 11 place Marcelin Berthelot, 75005 Paris, France}
\affiliation{Center for Computational Quantum Physics, Flatiron Institute,  
162 Fifth avenue, New York, NY 10010, USA} 
\affiliation{Centre de Physique Th\'eorique, Ecole Polytechnique, CNRS, Universit\'e Paris-Saclay, 91128 Palaiseau, France}
\affiliation{DQMP, Universit\'e de Gen\`eve, 24 quai Ernest Ansermet, CH-1211 Gen\`eve, Suisse}
\author{Michel Ferrero}
\affiliation{Centre de Physique Th\'eorique, Ecole Polytechnique, CNRS, Universit\'e Paris-Saclay, 91128 Palaiseau, France}
\affiliation{Coll\`ege de France, 11 place Marcelin Berthelot, 75005 Paris, France}

\begin{abstract}
One of the distinctive features of hole-doped cuprate superconductors 
is the onset of a `pseudogap' below a temperature $T^*$.
Recent experiments suggest that there may be a connection between 
the existence of the pseudogap and the topology of the Fermi surface.
Here, we address this issue by studying the two-dimensional Hubbard model 
with two distinct numerical methods.  
We find that the pseudogap only exists when the Fermi surface is hole-like and that,  
for a broad range of parameters, its opening is concomitant with a Fermi surface topology 
change from electron- to hole-like. 
We identify a common link between these observations:
the pole-like feature of the electronic self-energy
associated with the formation of the pseudogap is found to also control the degree of particle-hole asymmetry, and hence the Fermi surface topology transition.
We interpret our results in the framework of an SU(2) gauge theory of fluctuating 
antiferromagnetism. We show that a mean-field treatment of this theory in a metallic state with U(1) 
topological order provides an explanation of this pole-like feature, and a good description of 
our numerical results. 
We discuss the relevance of our results to experiments on cuprates 

\end{abstract}

\title{Pseudogap and Fermi surface topology in the two-dimensional Hubbard model}
\maketitle

\section{Introduction}

A very debated topic in the physics of high-temperature superconductors is the
nature of the 'pseudogap'~\cite{norman2005pseudogap,alloul} in their phase
diagram. Below a temperature $T^*(p)$ which is a decreasing function of the hole-doping level 
$p$, a pseudogap develops, corresponding to a suppression of low-energy excitations 
apparent in many experimental probes. 
Extrapolated to zero-temperature, $T^*(p)$ defines a
critical hole doping $p^*$ above which the pseudogap disappears as doping is increased. 
Another important critical value of the doping, denoted here $p_\mathrm{FS}$, 
is that at which the Fermi surface 
topology changes from hole-like to electron-like, corresponding to a Lifshitz transition. 
Recent experiments on \bismtwolayer\,(Bi2212) have suggested that the pseudogap may be very sensitive 
to the Fermi surface (FS) topology and that $p^*\simeq p_\mathrm{FS}$ in this compound~\cite{benhabib2015,loret2017}. 
In a simultaneous and independent manner from the present theoretical work, 
Doiron-Leyraud {\it et al.\/}~\cite{doiron_2017} recently performed a systematic experimental study 
using hydrostatic pressure as a control parameter 
in the $\mathrm{La_{1.6-x}Nd_{0.4}Sr_xCuO_4}$ (Nd-LSCO) system,
and an unambiguous connection between FS topology and the pseudogap
was found.

In this work, we investigate this interplay by studying the two-dimensional Hubbard model.
In the weak-coupling scenarios of pseudogap physics, there is a natural connection between the FS topology 
and the coherence of low-energy quasiparticles. Indeed, for a hole-like FS, 
coherence is suppressed at the `hot spots' where the FS intersects the 
antiferromagnetic zone boundary. When the FS turns electron-like, increased 
quasiparticle coherence is restored all along the FS (see Appendix
for a more detailed analysis)~\cite{honerkamp2001,abanov2003,Lohneysen2007,tremblay2006pseudogap,metlitski2010,montiel2017}.
At stronger coupling, several methods~\cite{dagotto1992,preuss1997,lichtenstein2000,maier2005review,
civelli2005dynamical, macridin2006, kyung2006pseudogap, tremblay2006pseudogap,scalapino2007,haule2007, ferrero2009,gull2009momentum,
gull2010, sordi2011, sordi2012, gunnarsson2015fluct,fratino2016, weiwu}
have established that the Hubbard model 
displays a pseudogap which originates from antiferromagnetic correlations. 
These correlations become short-range as the coupling strength or doping level are increased, as found in experiments~\cite{kastner1998}. 
The FS topology, on the other hand, is an issue which has to do with low-energy, 
long-distance physics. 
Hence, it is an intriguing and fundamentally important question to understand 
how the short-range correlations responsible for the pseudogap 
can be sensitive to FS changes.

Here, we study the Hubbard model for a broad range of parameters, and analyze the pseudogap and Fermi surface topology, and their interplay. 
We show that, at strong coupling, interactions can strongly modify the Fermi surface, making it more hole-like 
as compared to its non-interacting shape~\cite{maier2002angle,civelli2005dynamical,chen2012lifshitz,sakai2009evolution,Tocchio2012}.
We find that a pseudogap only exists when the FS is hole-like, so that $p^* \leq p_\mathrm{FS}$.  
We  identify an extended parameter regime in which these two critical doping levels are very close to one another: $p^* \simeq p_\mathrm{FS}$, so that  the FS turns electron-like only  when the pseudogap collapses. 
Moreover
we show that, when considering the 
relation between the pseudogap and FS topology, hole-doped cuprates can be 
separated into two families: materials for which $p^* \simeq p_{FS}$ and materials which 
have $p^* < p_{FS}$. These two families differ mostly by the relative magnitude of the 
next nearest-neighbor hopping.
These findings are shown to be consistent with a large body of experiments on cuprates.

We reveal that a common link between these observations is the pole-like 
feature~\cite{maier2002angle,stanescu2006fermi,berthod2006,gull2009momentum,sakai2009evolution,gull2010,lin2010} 
displayed by the electronic self-energy at the antinodal point, $\vec{k}=(\pi,0)$. 
The large imaginary part of the antinodal self-energy associated with this pole is responsible for the pseudogap, 
while the large particle-hole asymmetry associated with its real part controls the interaction-induced deformation of the Fermi surface and 
the location of the Fermi surface topology transition.
We investigate the evolution of this particle-hole asymmetry as a function of 
doping and nearest-neighbor hopping $t^\prime$, and show that the line in $(p,t^\prime)$ 
space where particle-hole symmetry is approximately obeyed at low energy is pushed, at strong coupling, 
to very low values of $p$ and very negative values of $t^\prime$. 
This is in stark contrast to the results of weak-coupling theories where this line is close to the Lifshitz transition 
of the non-interacting system. This also explains why interactions drive the Fermi surface 
more hole-like for hole-doping.

In order to understand these results from a more analytic standpoint, we
consider a recently developed SU(2) gauge theory of fluctuating
antiferromagnetic order \cite{SMQX09,CSS17}; additional results on the SU(2)
gauge theory appear in a companion paper, Ref.~\onlinecite{OurPreprint2}.  We
focus on a metallic phase of this theory, characterized by U(1) topological
order, which does not break spin or translational symmetries. We show that a
mean-field treatment of this gauge theory provides a good description of our
numerical results. In particular, the self-energy of the charge-carrying field
(chargon) in this theory displays a pole which provides an explanation for the
quasi-pole of the physical electron self-energy. The latter is calculated and
compares well to our numerical results, as do the trends in the evolution of
the pseudogap and  particle-hole asymmetry as a function of $p$ and $t^\prime$.

This paper is organized as follows. In Sec. II, we briefly introduce the model and the 
numerical methods  used in this article.
In Sec. III, we study the interplay between the pseudogap and FS topology 
and analyze the mechanisms controlling this interplay. 
The comparison and interpretation of our results in terms of the SU(2) gauge 
theory is presented at the end of this section.
In Sec. IV we discuss the relevance of our results to experiments on hole-doped cuprates. 
Sec.V provides a conclusion and outlook. 
Finally, details about the employed methods and various supporting materials 
can be found in the Appendices.

\section{Model and method}

We consider the Hubbard model defined by the Hamiltonian: 
\begin{equation}
 \mathcal{H} = -\,\sum_{ij,\sigma} t_{ij} c^{\dagger}_{i,\sigma} c^\pdagger_{j,\sigma}
    + U \sum_{i} n_{i\uparrow} n_{i\downarrow} - \mu \sum_{i,\sigma} n_{i\sigma},
\label{eq:Hubbard_Hamiltonian}    
\end{equation}
where $U$ is the onsite Coulomb repulsion and $\mu$ the chemical potential.
The hopping amplitudes $t_{ij}$'s are chosen to be non-zero between nearest-neighbor sites ($t_{ij}=t$) 
and next-nearest-neighbor ones ($t_{ij}=t'$). These hopping amplitudes define a non-interacting
dispersion relation $\epsilon_\bk=-2t(\cos k_x + \cos k_y) -4t'\cos k_x\cos k_y$. 
In the following, $t=1$ will be our unit of energy.
We solve this model using two distinct methods: the dynamical cluster approximation
(DCA~\cite{maier2005review}) and determinant quantum Monte Carlo 
(DQMC~\cite{bss1981}), see the Appendix for details.
Cluster extensions of dynamical mean-field theory 
(DMFT)  have shown that the Hubbard
model is able to capture many features of cuprate superconductors, such as the
superconducting dome and the 
pseudogap~\cite{lichtenstein2000,jarrell2001,maier2005review,
civelli2005dynamical, macridin2006, kyung2006pseudogap, tremblay2006pseudogap,haule2007,ferrero2009,werner2009,gull2009momentum,sakai2009evolution,
gull2010,gull2013,gunnarsson2015fluct}.
They have also established that the pseudogap originates from antiferromagnetic correlations,
which become short-range as the coupling strength or doping level are increased. 
This was also recently corroborated by exact diagrammatic Monte Carlo simulations~\cite{weiwu}. 
While
cluster extensions of DMFT have shown that hole doping can drive a
Lifshitz transition~\cite{maier2002angle,sakai2009evolution,chen2012lifshitz}
no general relationship between the pseudogap  and FS topology has been established.
We therefore carry out a systematic study for a broad range of parameters in order to investigate this issue.

\section{Results}

\subsection{Pseudogap and Fermi surface topology}

\begin{figure*}[!ht]
  \begin{center}
    \includegraphics[scale=1.0]{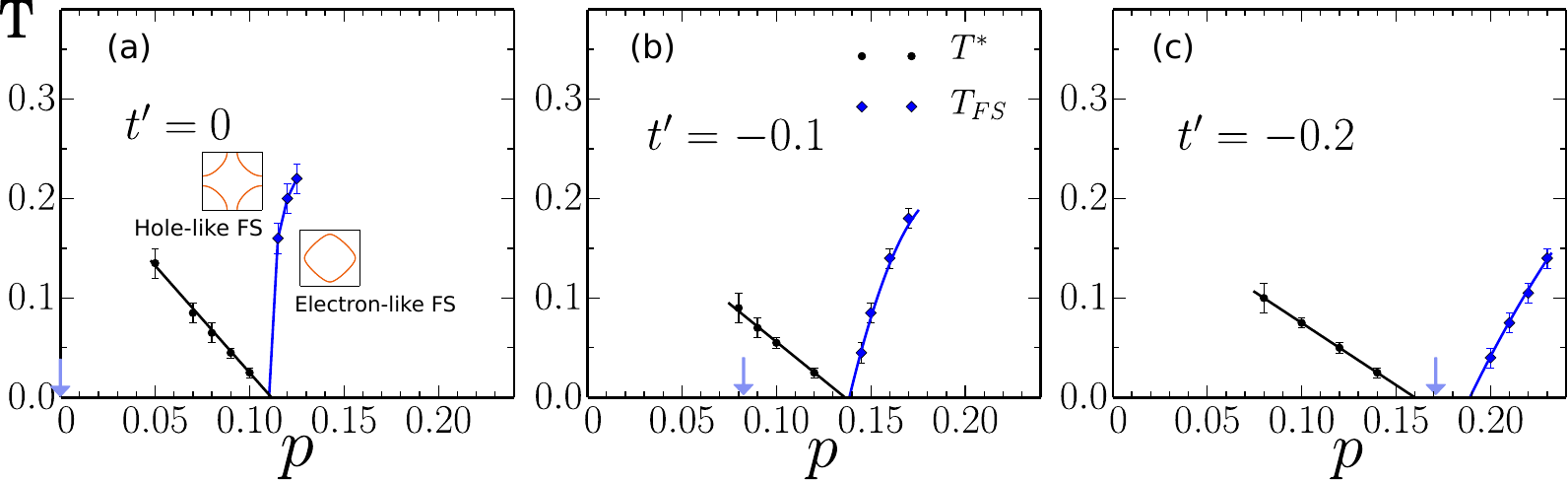}
  \end{center}
  \vspace{-0.4cm}
  \caption{\label{fig:tempvsdoping}
    \textbf{Pseudogap and Lifshitz transition temperatures.}
    Evolution of the pseudogap onset temperature $T^*$ (black) and
    the Lifshitz transition temperature $T_\mathrm{FS}$ (blue) as a function of the hole doping $p$, 
    for several values of $t'$ and $U=7$. The finite temperature data points are extrapolated
    to zero temperature and yield two critical dopings $p^*$ and
    $p_\mathrm{FS}$. It is apparent that $p^* \simeq p_\mathrm{FS}$ for $t'=0$ and $t'=-0.1$, while $p^* < p_\mathrm{FS}$ for $t'=-0.2$. 
    The solid lines are linear (for $T^{*}$) and quadratic (for $T_\mathrm{FS}$) least squares fits to the data points, except the $T_\mathrm{FS}$ line of $t' = 0$ where
     $T_\mathrm{FS}$ collapses to zero close to $p^*$ . 
    Error bars estimate all uncertainties in finding $T^*$ and $T_\mathrm{FS}$ with DCA
    (see also Appendix).
    Note that the change of topology of the Fermi surface for the interacting system occurs at a larger doping than 
    that of the non-interacting system 
    (indicated by a light-blue arrow).
    }
\end{figure*}

In Fig.~\ref{fig:tempvsdoping}, we display the pseudogap onset temperature $T^*(p)$, and the
temperature $T_\mathrm{FS}(p)$ at which the Fermi surface changes its topology, as a
function of doping level $p$, for several values of the next-nearest-neighbor hopping $t'$. 
$T^*$ is identified as follows: we calculate the zero-frequency extrapolated value of the
spectral function at the antinodal point ($\pi,0$);
we find that its temperature dependence displays a maximum which we identify 
as $T^*$. Below this scale, the antinodal spectral intensity decreases, signaling the opening of a pseudogap. 
$T_\mathrm{FS}$ is identified as the temperature where the Fermi surface crosses the $(\pi,0)$ point, 
and turns from hole-like to electron-like as temperature decreases (see below). 
Note that our definition of a Fermi surface is a pragmatic one: strictly speaking a Fermi
surface only exists at zero temperature. At finite temperatures, we define the
Fermi surface as the surface in momentum-space corresponding to the maximum of the spectral
intensity as it would be observed, {\it e.g.\/} in an angle-resolved photoemission (ARPES) experiment~\cite{shen2003}.

When extrapolated to zero temperature, these data
define two critical doping levels: 
$p^*$ such that the pseudogap disappears for $p > p^*$, and $p_\mathrm{FS}$ that marks the transition from a hole-like
FS ($p<p_\mathrm{FS}$) to an electron-like FS ($p>p_\mathrm{FS}$). 
Strikingly, the two curves in Fig.~\ref{fig:tempvsdoping} suggest that the pseudogap can only exist 
when the Fermi surface is hole-like, i.e. that $p^*\leq p_\mathrm{FS}$.
It appears that for values of $t' \ge -0.1$ both transitions happen at the same doping $p^* = p_\mathrm{FS}$ within our error bars. 
For more negative values of $t'$ the Fermi surface first becomes hole-like as $p$ is reduced, 
and the pseudogap opens at a lower doping, i.e. $p^* < p_\mathrm{FS}$.
We never observe a pseudogap with an electron-like Fermi surface, 
which would correspond to $p^* > p_\mathrm{FS}$.

\begin{figure}[!b]
  \begin{center}
    \includegraphics[scale=0.45]{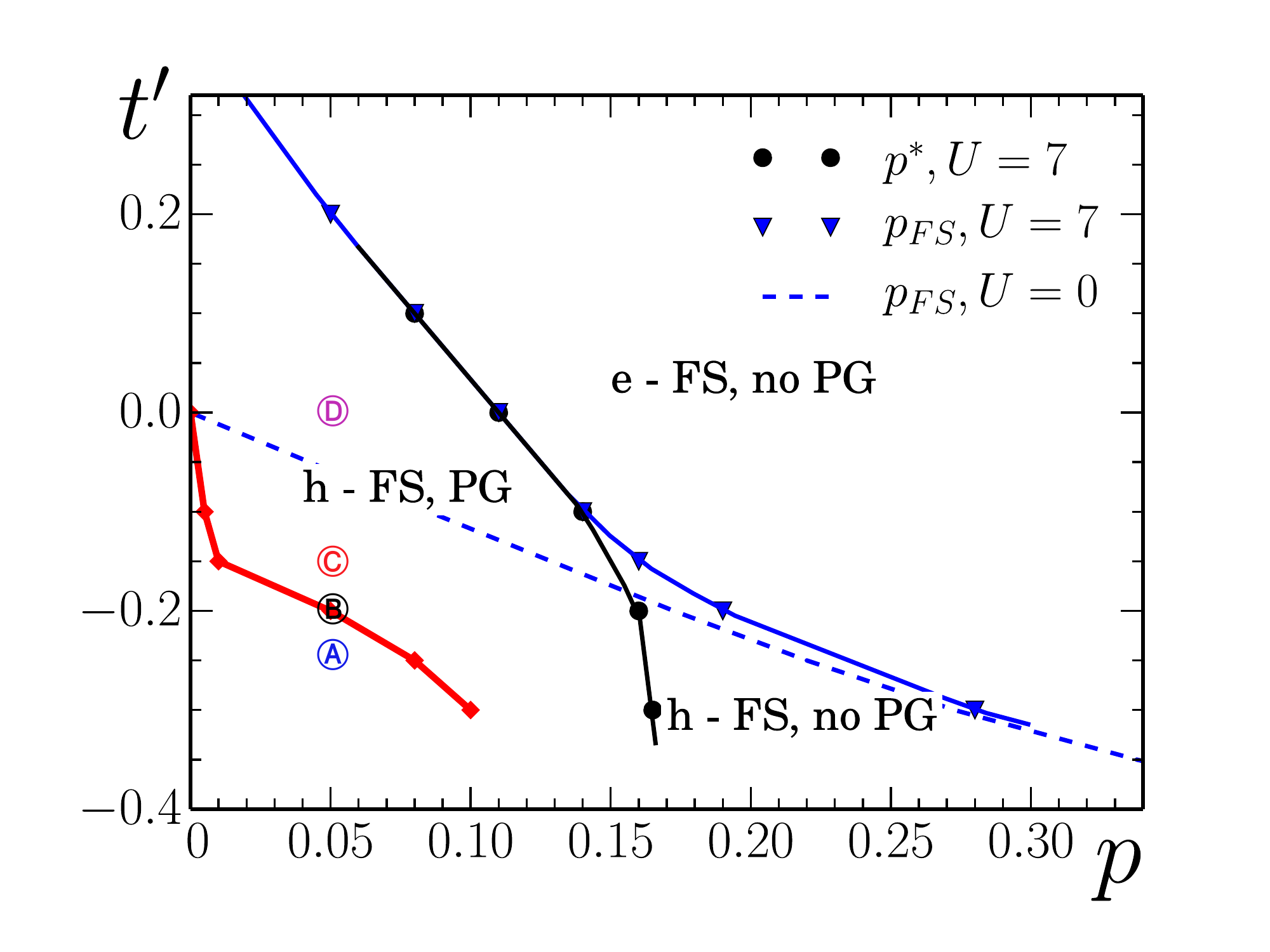}
  \end{center}
  \vspace{-0.6cm}
  \caption{\label{fig:phasediag}  
    \textbf{Zero-temperature 
    Fermi surface topology and pseudogap in the $p-t'$ plane.}
    The black line separates a region with no pseudogap (no PG) from a region
    where a pseudogap exists (PG). The blue line indicates where the
    interacting Fermi surface changes
    its topology from electron-like (e-FS) to hole-like (h-FS). The dashed blue line
    signals the same transition in the non-interacting case.
    The red curve locates the change in particle-hole asymmetry at the antinode: above the 
    red-line the real-part of the self-energy modifies the FS towards a more hole-like shape. 
    On the red line, the self-energy pole crosses zero-energy and approximate particle-hole symmetry is restored, 
    corresponding also to a maximum of the low-energy scattering rate as $t'$ is varied for fixed $p$.
    Points A-D label a set of parameters which are discussed further in Fig.~\ref{fig:scattering}. 
}
\end{figure}

This can be documented further by repeating this analysis for several 
doping levels $p$ and $t'$ values. 
The resulting map in the $(p,t')$ parameter space is displayed in Fig.~\ref{fig:phasediag}. 
A first observation is that the topological transition of the FS (blue line) that separates the regions
with hole-like and electron-like Fermi surfaces is strongly renormalized with
respect to its non-interacting ($U=0$) location (dashed line in Fig.~\ref{fig:phasediag} and 
arrows in Fig.~\ref{fig:tempvsdoping}).  
The black line defines the onset of the pseudogap. 
These lines define three regions: at large doping above the blue line, 
the FS is electron-like and no pseudogap is present. 
In the intermediate region between the two lines, the FS is hole-like but 
without a pseudogap. The topological transition and pseudogap opening 
coincide for a range of $t'$, while for more negative $t'$ the two lines split apart 
and, as doping is reduced, the pseudogap only opens after the FS has already turned 
hole-like at higher doping level ($p^*<p_\mathrm{FS}$).
The pseudogap and FS topology transition lines are  dependent 
on the value of $U$. 
As detailed in the Appendix, a larger value of $U$ yields 
a more extended regime of parameters 
for which $p^*\simeq p_\mathrm{FS}$, with the `branching point' where the two lines merge moving towards more negative values of $t'$ and larger doping level. 
This observation is important when comparing to experimental observations (see below).

\subsection{Change of Fermi surface topology due to correlation effects}

\begin{figure}
  \begin{center}
     \includegraphics[scale=0.5]{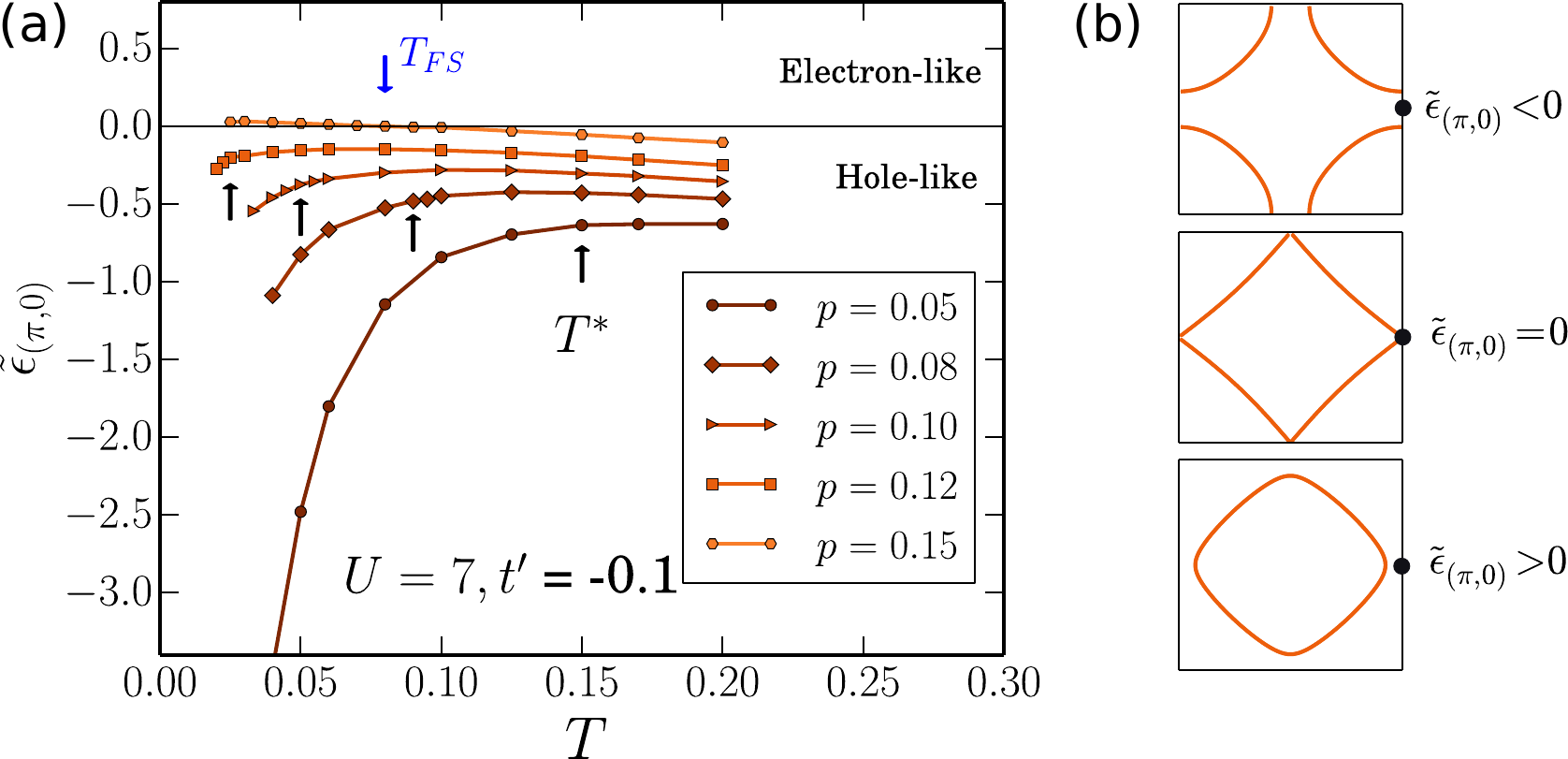}
  \end{center}
  \vspace{-0.4cm}
  \caption{\label{fig:ek}
    \textbf{Antinodal quasiparticle dispersion and Fermi surface topology} \textbf{(a)}  Antinodal quasiparticle energy   
    $\tilde\epsilon_{(\pi,0)}$ for different doping levels, as a
    function of temperature. The pseudogap onset temperature $T^*$ and the
    Lifshitz transition temperature $T_\mathrm{FS}$ are indicated by arrows.
    Below the pseudogap temperature, $\tilde\epsilon_{(\pi,0)}$ rapidly
    becomes very negative, driving the FS hole-like. Only when no pseudogap is present (here for $p > 0.12$) 
    does $\tilde\epsilon_{(\pi,0)}$ increase at low temperature and eventually becomes positive to yield an electron-like Fermi surface.
    \textbf{(b)} Illustration of the relation between the sign of $\tilde\epsilon_{(\pi,0)}$ and the Fermi surface topology.
  }
\end{figure}

The Fermi surface topology at the antinode is controlled by the renormalized
quasiparticle energy
\begin{eqnarray}\nonumber
  \tilde\epsilon_{(\pi,0)}&=&\epsilon_{(\pi,0)} - \mu + \mathrm{Re} \Sigma_{(\pi,0)}(\omega=0)\\ 
 &=&4t^\prime - \mu + \mathrm{Re} \Sigma_{(\pi,0)}(\omega=0)
  \label{eq:ANdispersion}
\end{eqnarray}
For negative values of $\tilde\epsilon_{(\pi,0)}$ the Fermi surface is
hole-like, while it is electron-like for $\tilde\epsilon_{(\pi,0)} > 0$. 
In order to gain insight in the
mechanisms driving the Lifshitz transition, Fig.~\ref{fig:ek} displays
$\tilde\epsilon_{(\pi,0)}$ as a function of temperature for various doping levels, with 
arrows indicating $T^*$ and $T_\mathrm{FS}$. Interestingly, even at the highest
temperature $T=0.2$ displayed there, $\tilde\epsilon_{(\pi,0)}$ is negative for all
doping levels, yielding a hole-like Fermi surface while the non-interacting Fermi
surface would be electron-like for $p\gtrsim 9\%$. 
This temperature is above the pseudogap temperature $T^*$, and hence  
the renormalization of the FS would be visible on a full Fermi
surface in an ARPES experiment. In this high-temperature range, 
only local correlations are responsible for this effect, as already
captured in a single-site DMFT calculation 
(see Fig.~\ref{ekdmft} in the Appendix).
As temperature is decreased, $\tilde\epsilon_{(\pi,0)}$ first increases slightly but then suddenly drops to
very negative values, pushing the Fermi surface to be very hole-like at low
temperatures. This starts happening just above the pseudogap temperature and
both effects can be traced back to non-local electronic correlations.  For this
value of $t' = -0.1$ the connection between the disappearance of the pseudogap
and the recovery of an electron-like surface is clear. Indeed, when no
pseudogap is present as e.g. for $p=0.15$, $\tilde\epsilon_{(\pi,0)}$ 
keeps on increasing and crosses zero, and an electron-like FS is recovered at low-$T$. 

\subsection{Particle-hole asymmetry and pole-like structure in the self-energy}

\begin{figure*}
  \begin{center}
     \includegraphics[scale=0.8]{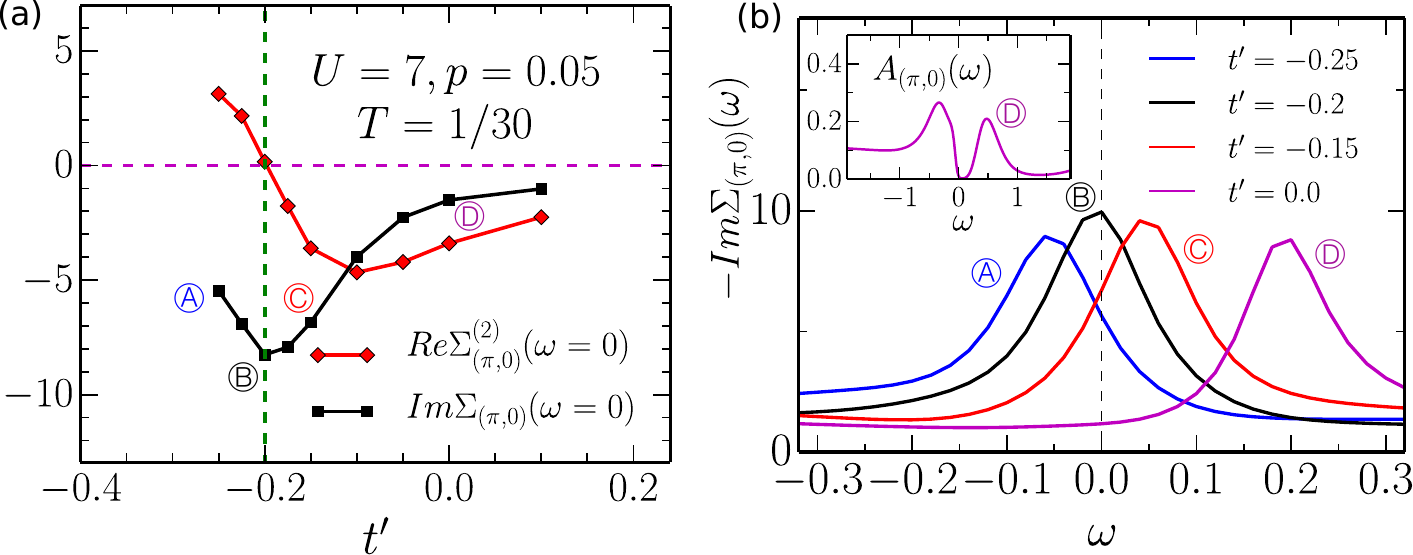}
  \end{center}
  \vspace{-0.4cm}
  \caption{\label{fig:scattering}
    \textbf{Evolution of the antinodal self-energy at fixed doping $p=0.05$, as a function of $t'$.}
    (a): Real (Hartree subtracted, see text) and imaginary parts of the antinodal self-energy at $\omega=0$.
    The real part vanishes where the imaginary part is maximum, corresponding to 
    a particle-hole symmetric low-energy $\mathrm{Im}\Sigma_{(\pi,0)}(\omega)$.
    (b): Real-frequency scattering rate $\mathrm{Im}\Sigma_{(\pi,0)}(\omega)$ obtained
    from the Maximum Entropy method for different values of $t'$. It displays
    a pole-like feature that crosses zero at $t' \simeq -0.2$ (point B) where the
    low-energy scattering is maximum. When the pole is on the positive
    energy side, it induces a negative real part of the self-energy (through
    the Kramers-Kronig relation) that drives the Fermi surface more
    hole-like. 
    Inset: antinodal spectral function at point D at $T=1/30$.    
    See Fig.~\ref{fig:phasediag} for the locations of points $A-D$ in the $(p,t')$ plane.
    }
\end{figure*}

From the definition of $\tilde\epsilon_{(\pi,0)}$ it is clear that it is the real part of the self-energy at the antinode 
that drives the renormalization of the FS. 
In Fig.~\ref{fig:scattering}a, we consider a fixed doping level $p=5\%$ and display 
$\mathrm{Re}\Sigma^{(2)}_{(\pi,0)}(\omega=0)$ as a function of $t'$, 
in which $\Sigma^{(2)}\equiv \Sigma-Up/2$ is the self-energy from which the Hartree (infinite frequency) 
contribution has been subtracted out.
It is seen that $\mathrm{Re}\Sigma^{(2)}_{(\pi,0)}(\omega=0)$  changes sign around $t' \simeq -0.2$ 
and becomes negative and fairly large for larger values of $t'$. This pushes the Fermi surface topology
transition to higher values of $t'$: for 5\% doping it remains hole-like up
to $t' \simeq +0.2$ whereas the Lifshitz transition of the non-interacting system 
occurs at $t' \simeq -0.05$ (see also Fig.~\ref{fig:phasediag}). 

The real part of the self-energy is related to its imaginary part through the Kramers-Kronig relation
\begin{equation}
\mathrm{Re}\Sigma^{(2)}_\bk(\omega = 0) = \frac{1}{\pi}
    \int_{0^+}^{\infty} \frac{\mathrm{Im}\Sigma_\bk(\omega') - \mathrm{Im}\Sigma_\bk(-\omega')}{\omega'} d\omega'.
\end{equation}
It is therefore instructive to analyze the behavior of $\mathrm{Im}\Sigma_{(\pi,0)}(\omega)$ (Fig.~\ref{fig:scattering}b) 
for several values of $t'$ (as indicated by the points A, B, C and D on Fig.~\ref{fig:phasediag}) corresponding to positive,
vanishing and negative values of $\mathrm{Re}\Sigma^{(2)}_{(\pi,0)}(\omega=0)$.
In all four cases, the imaginary part of the self-energy displays a prominent 
peak, corresponding to a pole-like feature of the self-energy. For $t'=-0.2$ (point B), 
this peak is centered at $\omega=0$. Because it is particle-hole symmetric, 
it leads to a vanishing real part of the self-energy (see Fig.~\ref{fig:scattering}a).
For values of $t'$ just  below and above -0.2 (points A and C), the
peak in $\mathrm{Im}\Sigma_{(\pi,0)}(\omega)$ shifts to negative (resp.~positive)
values of $\omega$. It has become particle-hole asymmetric and induces a
positive (resp.~negative) real part of the self-energy. There is therefore a
direct connection between the existence of a large particle-hole asymmetric
peak in the imaginary part of the self-energy and the renormalization of the
Fermi surface to a more hole-like topology. Note that the largest value of the
low-frequency scattering rate as $t'$ is varied is found when  
$\mathrm{Im}\Sigma_{(\pi,0)}(\omega)$ is particle-hole symmetric (e.g. point B in Fig.~\ref{fig:scattering}):
this defines the location of the red line 
in Fig.~\ref{fig:phasediag} (see also the Appendix). Anywhere above this
line, the self-energy is particle-hole asymmetric and drives the
Fermi surface topology transition to larger doping $p$ as compared to the non-interacting case.
Note that the system becomes very incoherent below the red line, 
at more negative values of $t'$ and small doping. 
The precise nature of the Fermi surface in this region, 
and its possible reconstruction, is difficult to assess with the methods employed here.

This pole-like feature in the self-energy is also responsible for opening the pseudogap,  
as clearly seen from the inset of Fig.~\ref{fig:scattering}b which displays 
the antinodal spectral function: the minimum of the spectral intensity is found to 
coincide with the frequency of the quasi-pole, where $\mathrm{Im}\Sigma_{(\pi,0)}(\omega)$ 
is largest.

\begin{figure*}
  \begin{center}
     \includegraphics[scale=1]{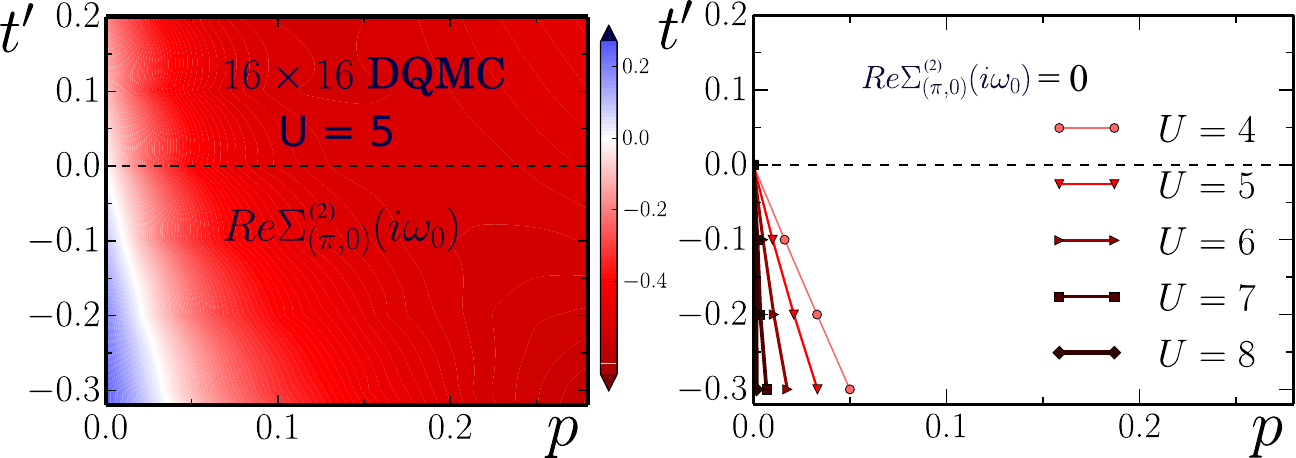}
  \end{center}
  \vspace{-0.4cm}
  \caption{\label{fig:qmc}
    \textbf{Particle-hole asymmetry from determinant quantum Monte
    Carlo (see methods).} 
    (a): Real part of the self-energy in the $p-t'$ plane for $U=5$. 
    For a broad region (indicated in red), $\mathrm{Re}\Sigma^{(2)}_{(\pi,0)}$ is
    negative, hence driving the Fermi surface more hole-like, 
    in agreement with our DCA results. 
    (b): The line where $\mathrm{Re}\Sigma^{(2)}_{(\pi,0)}=0$ and where the 
    antinodal scattering rate is largest is indicated for 
    different values of $U$ (to be compared to the red line in Fig.~   \ref{fig:phasediag}). 
    As $U$ is increased the region where the FS is driven hole-like becomes 
    larger.
  }
\end{figure*}

\subsection{Fermi surface topology: numerically exact DQMC results}

These results have been obtained using the DCA approximation with an
$8$-site cluster (see Appendix~\ref{app:DCA}). 
We also cross-checked these results with a different and independent method: 
numerically exact determinant quantum Monte Carlo (DQMC)~\cite{bss1981} at $T=1/3$. 
The result is displayed in Fig.~\ref{fig:qmc} (left panel) and clearly shows 
that the antinodal self-energy drives the Fermi surface hole-like over a broad region of the
$(p,t')$ plane, in agreement with our DCA calculations.  One can again observe a
line where $\mathrm{Re}\Sigma^{(2)}_{(\pi,0)}(\omega)$ vanishes, mapped out for
several values of $U$ in the right panel.  This line compares with the red line
of Fig.~\ref{fig:phasediag}, and moves closer to half-filling as $U$ is increased (see also Fig.~\ref{u7u75}) and 
towards the non-interacting Lifshitz transition line as $U$ is reduced. 

\subsection{SU(2) gauge theory}
\label{sec:gauge_theory}

Recent numerical work, using a `fluctuation diagnostics' analysis of the contributions to the 
electronic self-energy in both the DCA~\cite{gunnarsson2015fluct} and lattice diagrammatic Monte-Carlo~\cite{weiwu} 
approaches have established that the pseudogap is associated with the onset of short-range 
antiferromagnetic (AF) correlations. 
On the analytical side, an SU(2) gauge theory approach has been introduced \cite{SMQX09,CSS17,OurPreprint2}
to deal with states in which AF long-range order is destroyed by orientational 
fluctuations of the order parameter. 
It is thus very natural to attempt to interpret our numerical results in this framework and compare 
them to a mean-field treatment of this gauge theory. 
 
This approach is based on the following representation of the physical electron fields on each lattice site $i$: 
\begin{equation}
\left(\begin{matrix}
c_{i\uparrow} \\ c_{i\downarrow}
\end{matrix} 
\right)\,=\, 
R_i\,
\left(\begin{matrix}
\psi_{i +} \\ \psi_{i -}
\end{matrix} 
\right)
\end{equation}
In this expression, $\psi_{\pm}$ are `chargons' - fermions which carry charge but no spin quantum numbers 
and $R_i$'s are $2\times 2$ unitary matrix fields, the bosonic spinons ($R_iR_i^\dagger=R_i^\dagger R_i = 1$). 
The $R_i$ matrix can be thought of as defining the local reference frame associated with the local AF 
order (for early work promoting the local reference frame to a dynamical variable, see Refs.~\onlinecite{SS88,Schulz90,Schrieffer2004}).
This representation has a local gauge invariance corresponding to 
$R_i\rightarrow R_i V_i^\dagger, \psi_i\rightarrow V_i \psi_i$, with $V_i$ an SU(2) matrix. 
The Hubbard interaction can be decoupled using a vector field $\vec{\Phi}_i$ conjugate to the local 
spin-density $c^\dagger_{i\alpha}\vec{\sigma}_{\alpha\beta}c_{i\beta}/2$, and a vector `Higgs field' is 
introduced such that: 
\begin{equation}
\vec{\sigma}\cdot\vec{H}_i\,=\,R^\dagger_i \vec{\sigma} R^\pdagger_i\cdot \vec{\Phi}_i \,.
\end{equation} 
This identifies the Higgs field, $\vec{H}_i$, as the local antiferromagnetic moment in the rotated reference frame.
Note that $\vec{H}_i$, which transforms under the adjoint of the gauge SU(2), 
does not carry any spin since it is invariant under a global spin rotation. 

We can now consider Higgs phases in which $\langle \vec{H}_i \rangle \neq 0$ but $\langle R_i \rangle =0$. 
Because of the latter, such phases do not display long-range AF order, which has been destroyed by 
orientational fluctuations. However, $\langle \vec{H}_i \rangle \neq 0$ signals that the  
local order has a non-zero amplitude. 
A non-zero $\langle \vec{H}_i \rangle$ also implies that such a phase has {\it topological order}, corresponding to 
different possible residual gauge groups once the SU(2) gauge symmetry has been spontaneously broken by the Higgs 
condensate \cite{FradkinShenker,NRSS91,Wen91,Bais92}. 
There are different possible mean-field solutions for the Higgs condensate, corresponding to different topological 
orders and different broken discrete symmetries \cite{CSS17}. 
Here we shall focus on the simplest one with U(1) 
topological order which preserves all space group, time-reversal, and spin rotations symmetries; this corresponds to the following configuration of the Higgs field (which resembles AF order):
\begin{equation}
\langle \vec{H}_i\rangle \,=\, \left(0,0,H_0\,e^{i\vec{Q}\cdot\vec{R}_i} \right) \,,
\end{equation}
in which $H_0$ is the Higgs field amplitude and $\vec{Q}=(\pi,\pi)$. 

Solving the gauge theory at the mean-field level, the Green's function and self-energy of 
the chargon field is easily calculated. 
Because the chargon field `sees' an antiferromagnetic environment, it is identical to the 
expression obtained for an antiferromagnetic spin-density wave\cite{OurPreprint2}. It thus has a matrix form which involves 
both components which are diagonal in momentum and off-diagonal components coupling $\vec{k}$ to $\vec{k+Q}$:
\begin{equation}
G_{\psi} (\omega,\vec{k})^{-1}\,=\,
\left( 
\begin{array}{cc}
\omega- \xi^\psi_{\vec{k}} & H_0 \\
H_0 & \omega- \xi^\psi_{\vec{k+Q}}
\end{array}
\right)
\end{equation}
Its momentum diagonal component reads 
 \begin{eqnarray}\nonumber
 G_{\psi} (\omega,\vec{k}) &=&\left[\omega- \xi^\psi_{\vec{k}}-\Sigma_{\psi}(\omega,\vec{k})\right]^{-1}\\
  \Sigma_{\psi}(\omega,\vec{k})&=&\frac{ H_0^2}{\omega -\xi^\psi_{\vec{k}+\vec{Q}}+i0^+} 
 \label{eq:GreensFunctionChargon}
 \end{eqnarray}
In this expression, $\xi^\psi_{\vec{k}}=-2Z_t t (\cos k_x+\cos k_y) -4Z_{t'} t^\prime \cos k_x\cos k_y -\mu$ is 
the renormalized dispersion of the chargons. 
A quantitative calculation of the renormalization factors $Z_t$ and 
$Z_{t'}$ requires a full solution of the mean-field equations. We found typical values 
$Z_{t} \sim 0.3$ and $Z_{t'} \sim 0.2$, weakly dependent on the doping level $p$ since the chemical potential mainly affects the 
chargon dispersion but not the spinon dispersion. 
Importantly,
the self-energy (\ref{eq:GreensFunctionChargon}) of the chargons 
has a pole at $\omega_{\vec{k}}=\xi^\psi_{\vec{k}+\vec{Q}}$. 
Hence the mean-field chargon Green's function has zeros: these zeros are located at zero energy on the Brillouin zone contour 
defined by $\xi^\psi_{\vec{k}+\vec{Q}}=0$, corresponding to a chargon `Luttinger surface'.
There are two bands of chargon excitations, corresponding to the solutions of 
$(\omega-\xi^\psi_{\vec{k}})(\omega-\xi^\psi_{\vec{k}+\vec{Q}})-H_0^2=0$. 
To summarize, a crucial aspect of this SU(2) gauge theory description is to have chargons whose 
dispersions are identical (at the mean-field level) to the excitations of a spin-density wave states, {\it despite 
the theory having no long-range order or broken symmetries} (i.e. the symmetry is restored by the fluctuations  
of the spinon fields).

At the mean-field level, in the phase associated with the configuration of the Higgs field considered here, 
the spinon excitations are gapped.
In order to obtain the physical electron Green's function, a convolution of the chargon and spinon 
Green's function over frequency and momentum must be performed: $G_c=G_R\star G_\psi$ and the physical 
electron self-energy can then be obtained from $\Sigma = \omega+\mu-\epsilon_{\vec{k}}-G_c^{-1}$ 
(with $\epsilon_{\vec{k}}$ the bare dispersion defined above). 
For the purpose of the present paper, a detailed discussion of the spinon dispersion and Green's function 
is not essential, see Appendix~\ref{app:convolution} and Ref.~\onlinecite{OurPreprint2} for details. 
It is sufficient here to emphasize the two following points. 
(i) The convolution mainly broadens the
pole structure of $G_\psi$ but the location in momentum and frequency of the
most singular structures of the physical self-energy are still those encoded in
the chargon self-energy given by~(\ref{eq:GreensFunctionChargon}). 
(ii) The convolution does bring an important effect however: in contrast to the imaginary part of the 
chargon self-energy, which is constant all along the Luttinger surface $\xi_{\bk+\bQ}^\psi=0$, 
the imaginary part of the physical electron self-energy obtained from the convolution of Green's 
functions has an imaginary part which is larger close to the antinodes than close to the nodes, 
see Fig.~\ref{fig:su2sigma} in Appendix~\ref{app:convolution}. Hence, the gauge theory manages to capture 
qualitative aspects of the nodal-antinodal dichotomy found in our DCA calculations.

The figure also shows that the peak frequency $\omega_p$ shifts from negative
to positive frequency as $t^\prime$ is increased. The inset of this figure
displays the corresponding spectral function at the antinode, which has a
pseudogap caused by the quasi-pole at $\omega_p$. Note that the pseudogap is
particle-hole asymmetric, as expected from the fact that it does not originate
from the particle-particle channel. These results are in excellent qualitative
agreement with the DCA calculations above (Fig.~\ref{fig:scattering}).
Note that, for the sake of comparison to the finite-temperature DCA results, the gauge theory calculations 
presented here are performed at a finite temperature larger than the spinon gap. How do gapless nodal excitations 
survive in the gauge theory description as temperature is lowered below this gap (e.g. by having bound-states of 
the chargons and spinon as in an FL* state~\cite{sss_FL*}) is an important question which 
is however beyond the scope of the present paper.

\begin{figure*}
  \begin{center}
     \includegraphics[scale=1.0]{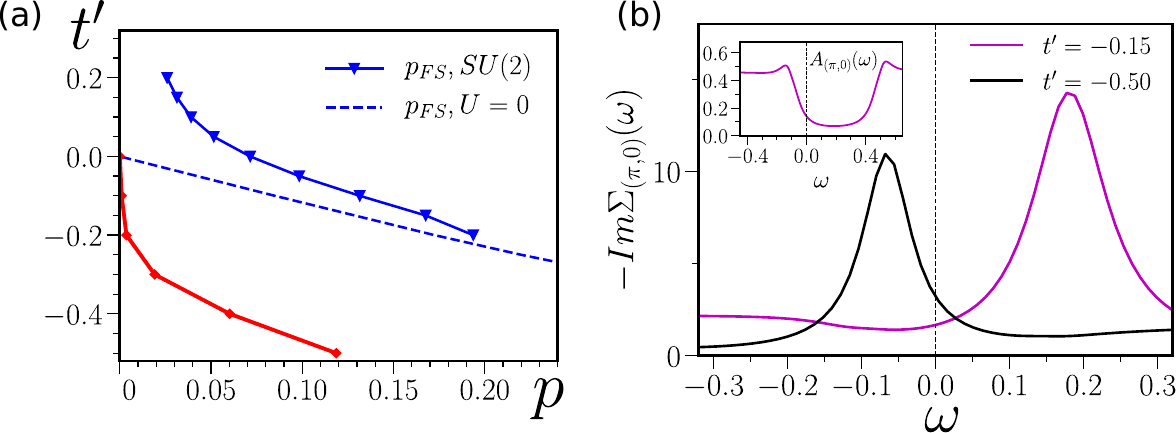}
  \end{center}
  \vspace{-0.4cm}
  \caption{
    \textbf{Pseudogap and Fermi surface topology within the SU(2) gauge theory at mean-field level.} 
    Color coding is identical to Fig.~\ref{fig:phasediag} (DCA results), to which this figure should be compared. 
    \textbf{(a)} Solid blue line: Lifshitz transition of the interacting Fermi surface. 
    Along the red line, the self-energy pole is at zero energy and approximate particle-hole symmetry is restored. 
    This also corresponds to the Lifshitz transition of the chargons.  
    \textbf{(b)} Electronic self-energy at the antinode for two different values of $t^{\prime}$. 
    The quasi-pole in the self-energy moves from negative to positive frequency as $t^{\prime}$ is increased, see Fig.~\ref{fig:scattering}. 
    The inset displays the antinodal spectral function for $t^{\prime}=-0.15$, emphasizing that its minimum coincides with the 
    position of the pole. 
    Here we assumed $H_0 = 0.3$, a spinon gap $\Delta = 0.01$, and $J=0.1$ (the nearest-neighbor coupling of the spin-wave
    fluctuations, see Ref.~\onlinecite{OurPreprint2}). 
    A broadening factor $\eta = 0.04$ is used to obtain smooth spectral functions.
    }
\label{fig:gauge}
\end{figure*}

In Fig.~\ref{fig:gauge}(a), we summarize important aspects of the mean-field analysis of the gauge theory~\cite{OurPreprint2}  
as a function of doping level $p$ and $t^\prime$. As in Fig.~\ref{fig:phasediag}, the blue line 
in this figure is the location of the Lifshitz transition of the physical electron FS from hole to electron-like 
(as defined by the change of sign of the renormalized antinodal dispersion, Eq.~\ref{eq:ANdispersion}) and the 
red line indicates where $\omega_p=0$ (i.e. where particle-hole symmetry is approximately restored at low energy). 
In good qualitative agreement with the DCA results displayed in Fig.~\ref{fig:phasediag}, one sees that the Lifshitz 
transition of the physical FS is pushed to much larger doping in comparison to that of the non-interacting system 
(dashed line), and that the location of the red line where the pole is close to zero energy is pushed to much smaller doping.
The latter approximately coincides with the Lifshitz transition of the chargon
Luttinger surface, given by $4Z_{t^\prime}t^\prime=\mu$.
Because the chemical potential $\mu$ of the interacting system
takes more negative values than the non-interacting one and also because $Z_{t^\prime}< 1$, the red
line is shifted to lower doping as $U$ increases, in agreement with the result of
Fig.~\ref{fig:qmc}. This clarifies why the pole is found at positive energies
for most values of $(p,t')$ and why the FS id driven hole-like in a wide region of the ($p,t^{\prime}$) plane.
A striking consequence of the presence of the pole is illustrated around the $t^{\prime} =0,p=0 $ point, corresponding to the 
half-filled Hubbard model with only nearest-neighbor hopping, in which the antinodal scattering 
must be particle-hole symmetric by symmetry. When the system is very slightly hole-doped away from $p=0$, both DCA and 
the mean-field gauge theory suggest that the particle-hole symmetric point rapidly shifts to very negative $t^{\prime}$. 
This is in striking contrast to weak-coupling theories in which approximate particle-hole symmetry at the antinode 
would be restored at the non-interacting Lifshitz transition (dashed line). 
We note that there are quantitative discrepancies in the location of these two lines between the numerical DCA 
results and the mean-field gauge theory results, which are predominantly due to the assumptions made on  
the renormalization parameters $Z_t$ and $Z_{t^\prime}$ entering the chargon dispersion and on the 
Higgs field amplitude $H_0$.   

Importantly, the mean-field analysis of the SU(2) gauge theory provides a physical understanding 
of the origin of the pseudogap and of the quasi-pole of the self-energy as being due to short-range antiferromagnetic correlations, 
long-range order being destroyed by orientational fluctuations. 
The quasi-pole is responsible for the pseudogap in the physical electron Green's function, while the
spinon (R) spectrum displays a gap. 
The chargons have a spectrum characteristic of an AF spin-density wave despite the absence of AF long-range order, 
and their self-energy has a sharp pole at mean-field level. The (red) line where the 
pole crosses zero energy, corresponding to an approximate restoration of particle-hole symmetry at low-energy, 
can be interpreted\cite{OurPreprint2} as the Lifshitz transition of the chargon Luttinger surface.

\section{Discussion and experimental relevance}

Our results establish that an asymmetric pole-like feature in the antinodal self-energy 
is responsible for both the pseudogap and for the renormalization and topological transition of the FS. 
We note that, in weak-coupling approaches such as spin-fluctuation theories (see Appendix~\ref{app:weak} for a detailed discussion)  
the self-energy becomes very large for $\omega=\epsilon_{\bk+(\pi,\pi)}-\mu$, provided that the antiferromagnetic 
correlation length $\xi$ is large enough and that $v_F/\xi<T$. 
As a result, hot spots form on the Fermi surface, at specific $\bk$-vectors defined by $\epsilon_\bk=\epsilon_{\bk+(\pi,\pi)}=\mu$, 
corresponding to the intersection of the antiferromagnetic Brillouin zone with the Fermi surface. 
Hence, in a weak coupling approach, the change of sign of the bare dispersion $\epsilon_{(\pi,0)}-\mu=0$ controls both the doping 
at which the hot spots reach the antinode and that where the Lifshitz transition occurs. 
As a result, the non-interacting FS transition line (blue dashed line in
Fig.~\ref{fig:phasediag}) controls at the same time the location of the Lifshitz transition, 
the symmetry of the self-energy and the suppression of spectral weight along the Fermi
surface. This is in stark contrast to our  strong-coupling results where these phenomena appear at 
distinct locations.
In particular, we have demonstrated that the line in $(p,t^\prime)$ parameter space 
where particle-hole symmetry is approximately obeyed 
at low energy is pushed, at strong coupling, to very low values of $p$ and very negative values of $t^\prime$,
see Fig.~\ref{fig:phasediag} and Fig.~\ref{fig:qmc} where this line is displayed in red. 
This is crucial in explaining why interactions drive the Fermi surface 
more hole-like for a wide range of $(p,t^\prime)$ where the non-interacting (or weak coupling) 
FS would actually be electron-like, and why the Lifshitz transition is pushed to larger values of $p$ in comparison to the 
non-interacting system.

In order to put our results in perspective, we note that the relation between a pole-like feature in the self-energy 
and the pseudogap, as well as the implications of the corresponding zeros of the Green's function for the reconstruction of the Fermi surface  
have been previously discussed in cluster extensions of dynamical 
mean-field theory~\cite{maier2002angle,stanescu2006fermi,berthod2006,gull2009momentum,sakai2009evolution,gull2010,lin2010} 
and in phenomenological theories such as YRZ~\cite{YRZ2006} or other approaches~\cite{sakai2016hidden,dave_zeros_2013}  
(see Ref.~\onlinecite{qi_subir} for a gauge-theory perspective on the YRZ phenomenology). 
The existence of a Lifshitz transition as the hole doping is increased was also discussed in some previous cluster DMFT 
or DCA studies~\cite{maier2002angle,sakai2009evolution,chen2012lifshitz}.
However, the role played by the particle-hole asymmetry associated with the self-energy pole in determining the FS topology, 
and the systematic dependence of this asymmetry on $(p, t^{\prime})$ were not unraveled and studied, and hence the key interplay between 
FS topology and the pseudogap was not previously revealed.

We now discuss the relevance of our results to experiments on hole-doped cuprates.
We first note that, indeed, a pseudogap is not found when the FS is electron-like and hence 
that the relation $p^*\leq p_\mathrm{FS}$ is apparently obeyed in all compounds. 
In the single-layer compound $\mathrm{La_{2-x}Sr_xCuO_4 }$ (LSCO), 
with a small value~\cite{pavarini2001} of $|t'/t|$, 
the in-plane resistivity in high magnetic fields~\cite{cooper2009anomalous} suggests
that $p^*\simeq 0.18$. Currently available ARPES 
experiments~\cite{chang2008anisotropic,yoshida2006systematic,park_LSCO_2013} 
allow to ascertain that $0.17 < p_\mathrm{FS} \lesssim 0.20$. 
In the Nd-LSCO compound, high-field transport~\cite{collignon2016}
finds $p^* \simeq 0.23$, while ARPES~\cite{matt2015} has  $0.20 < p_\mathrm{FS} < 0.24$.

In another single-layer compound $\mathrm{(Bi,Pb)_ 2(Sr,La)_2CuO_{6+\delta}}$
(Bi2201)~\cite{he2014fermi, zheng2005critical, kawasaki2010carrier, kondo_Bi2201_JESRP_2004}, it is
found that  $p^*\simeq p_\mathrm{FS}$.
An ARPES experiment on the bilayer Bi2212
material~\cite{kaminski2006change} has shown that the antibonding FS crosses
the antinode at $p_\mathrm{FS}\simeq 0.22$ and suggested that it may be connected to
the onset of the pseudogap. This was further confirmed in a recent electronic
Raman experiment~\cite{benhabib2015, loret2017} that found the pseudogap
end-point at $p^* \simeq 0.22$. Note that the Raman response is believed to be
predominantly sensitive to the antibonding band since it is close to a density
of states singularity~\cite{benhabib2015} and does not give information about
the possible existence of a pseudogap in the bonding band (which remains
hole-like for all dopings).
In compounds with larger values~\cite{pavarini2001} of $|t'/t|$, such as 
$\mathrm{YBa_2Cu_3O_{7-\delta}}$~\cite{hossain2008situ,badoux2016change},
$\mathrm{Tl_2Ba_2CuO_{6+\delta}}$~\cite{plate2005fermi,proust2002heat} or
$\mathrm{ HgBa_2CuO_{4+\delta}}$~\cite{tabis2014charge}, it is generally
believed that $p_\mathrm{FS}$ and $p^*$ are distinct with $p^* < p_\mathrm{FS}$. 
This is in qualitative agreement with our finding that the FS and pseudogap 
critical doping coincide for smaller values of $|t'/t|$, 
and are distinct for larger ones. 
Hence, we conclude on the basis of our results and experimental observations that 
there are two families of hole-doped cuprates: materials with smaller values 
of $|t'/t|$ for which the collapse of the pseudogap and change of FS topology coincide 
($p^* \simeq p_\mathrm{FS}$), and materials with larger values of $|t'/t|$ for which 
these are distinct phenomena ($p^* < p_\mathrm{FS}$).

Finally, a very recent study on Nd-LSCO using  hydrostatic pressure to tune the
band structure finds that both $p_\mathrm{FS}$ and $p^*$  decrease by the same
amount.~\cite{doiron_2017} This provides a compelling experimental
demonstration that $p^*$ cannot exceed $p_\mathrm{FS}$.

We finally comment on the predicted renormalization of the FS by strong
correlations. In view of Fig.~\ref{fig:phasediag}, the materials for which this
effect is expected to be strongest are the ones with smaller values of
$|t'/t|$, hence we turn to LSCO. We note that, in order to fit the ARPES FS
using a single-band tight binding model, the effective parameter $t^{\prime}$
has to be tuned systematically more negative (corresponding to a more negative
$\tilde\epsilon_{(\pi,0)}$) as doping is reduced, \textit{i.e.,} from
$t^{\prime}/t = -0.12$ for $p=0.3$ to $t^{\prime}/t = -0.2$ for $p=0.03
$~\cite{yoshida2006systematic}.  Moreover, electronic structure calculations
based on DFT-LDA yield $p_\mathrm{FS}\simeq 0.15$ while ARPES finds $0.17 < p_\mathrm{FS}
\lesssim 0.20$, as mentioned above.  These two observations suggest that
correlation effects indeed generally drive the FS more hole-like.

\section{Conclusion and outlook}

To conclude, we have investigated the interplay between the pseudogap and
the Fermi surface topology in the two-dimensional Hubbard model.
In the weak-coupling regime these issues are directly connected: 
hot-spots can only form when the FS is hole-like and intersects the 
antiferromagnetic zone boundary. 
At stronger coupling, the antiferromagnetic correlations responsible 
for the pseudogap become short-ranged, and it becomes a fundamental puzzle 
to understand whether there is any connection to FS topology, 
which is in essence long-distance physics. 
We provide an answer to this puzzle by showing that a common pole-like 
feature in the electronic self-energy controls both issues. 
This pole induces a large low-energy scattering rate responsible 
for the onset of the pseudogap, and its asymmetry leads to significant modifications 
of the Fermi surface with respect to its non-interacting shape and controls 
the location of the Lifshitz transition. 
As a consequence,
we find that the pseudogap only appears on hole-like Fermi surfaces, 
i.e.  $p^* \le p_\mathrm{FS}$ and that $p^* \simeq p_\mathrm{FS}$ 
for an extended range of doping levels and values of $t'$. 
These findings are in good agreement with available experimental data. 
We have also shown that our results can be interpreted in the framework of 
an SU(2) gauge theory of fluctuating antiferromagnetism with topological order. 
This provides an explanation for the origin of the pole in the self-energy
and establishes the connection between the pseudogap and the Fermi surface
topology through the chargon Luttinger surface.
This effort to bridge the gap between numerical results obtained within cluster 
extensions of DMFT and low-energy effective field theories is pursued and detailed 
in a companion publication \cite{OurPreprint2}. 

Let us emphasize that in most of the parameter range relevant to hole-doped cuprates, 
the self-energy pole is found at a positive energy. Hence, the strongest suppression of the antinodal 
spectral weight is predicted to occur 
at energies above the Fermi level, which is not directly accessible to ARPES experiments.
While a strong particle-hole asymmetry is indeed observed by STM~\cite{kohsaka2007,fischer2007}, 
this emphasizes again~\cite{sakai2013} the importance of developing momentum-resolved spectroscopies 
able to probe the `dark side' of the FS.
Finally, an outstanding question is to explore whether the topological order, associated with the 
pseudogap regime in the gauge theory description, can be revealed more directly in numerical studies of Hubbard-like models.

\begin{acknowledgments}

We are grateful to L.~Taillefer and N.~Doiron-Leyraud for sharing and
discussing their experimental data before publication. We also acknowledge
discussions with M.~Civelli, M.~Kim, G.~Kotliar, A.~J.~Millis,
O.~Parcollet, I.~Paul, A.~Sacuto and A.-M.S. Tremblay.  This work was
supported by the Simons Foundation Many-Electron Collaboration, the European
Research Council (project ERC-319286-`QMAC'), the Swiss National Supercomputing
Centre (CSCS, project s575), the NSF under Grant DMR-1664842 and MURI grant
W911NF-14-1-0003 from ARO.  The Flatiron Institute is supported by the Simons
Foundation (AG).  Research at Perimeter Institute (SS) is supported by the
Government of Canada through Industry Canada and by the Province of Ontario
through the Ministry of Research and Innovation.  MS acknowledges support from
the German National Academy of Sciences Leopoldina through grant LPDS 2016-12.
SS acknowledges support from Cenovus Energy at Perimeter Institute, and from
the Hanna Visiting Professor program at Stanford University.  AG and SS are
grateful for the stimulating atmosphere of the May 2017 Jouvence workshop.

\end{acknowledgments}

\appendix
\section{Methods}
\label{app:DCA}

Our results for the two-dimensional Hubbard model are obtained using two
methods: unbiased determinant quantum Monte Carlo (DQMC~\cite{bss1981}) and the
dynamical cluster approximation (DCA~\cite{maier2005review, hirschfye}), a
cluster extension of dynamical mean-field theory
(DMFT~\cite{georges1996review}) that captures the physics of short-range
spatial correlations. We perform DQMC on a $16\times 16$ lattice with periodic
boundary conditions at a temperature $T=1/3$.  Since the inverse temperature
$\beta = T^{-1} = 3$ is significantly smaller than the linear size of the
lattice $L=16$, the finite size effects are negligible in the DQMC calculation. The
imaginary time step was set to $\Delta \tau = 3/64$ which is small enough to
avoid artifacts due to the discretization errors. We use $5.12 \times 10^{5}$ Monte Carlo sweeps to
collect the data after $1000$ warmup sweeps.

\begin{figure}[!b]
  \begin{center}
    \includegraphics[scale=1.2]{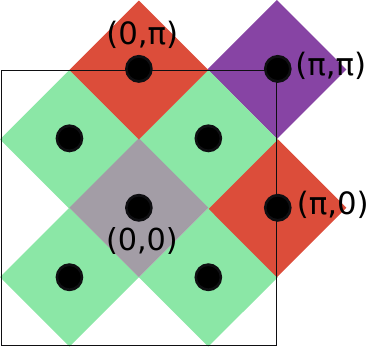}
  \end{center}
  \vspace{-0.4cm}
  \caption{The patches in momentum space of the eight-site DCA method. The self-energy is
  constant over each patch.
  \label{patch}}
\end{figure}

The DCA calculation is performed with an eight-site cluster.
In the DCA approach, the lattice
self-energy is approximated by a patchwise-constant self-energy $\Sigma_{\bK}$
in the Brillouin zone.  We solved the DCA equations with an eight-site
auxiliary quantum impurity cluster. In the geometry we used, the Brillouin
zone is divided in eight sectors where the self-energy is constant, as
shown in Fig.~\ref{patch}. Note that there are clearly distinct patches
for the antinodal and the nodal region of the Brillouin zone.

We use both the
the Hirsch-Fye~\cite{hirschfye} and the continuous-time quantum Monte
Carlo~\cite{rubtsov2005} method to solve the auxiliary cluster impurity
problem. A comparison of both methods shows that the imaginary-time step
$\Delta \tau  = 1/21$ used in the Hirsch-Fye solver is small enough, as shown
in the Appendix B.  We use 50 DMFT
iterations to get converged result and use $(2 \sim 10) \times 10^{6}$ Monte
Carlo sweeps at each iteration.  In order to have better statistics, the
results are averaged over the last few converged iterations. The typical
statistical error in the real part of the self-energy and the spectral
intensity at zero frequency is $\sim 1\%$.

We identify the pseudogap temperature $T^{*}$ as the maximum of the temperature
dependent spectral weight $A_{(\pi,0)}(\omega=0) \equiv -\frac{1}{\pi}
\mathrm{Im}G_{(\pi,0)}(\omega = 0)$. It is obtained from a linear extrapolation to
zero frequency of $-\mathrm{Im} G_{(\pi,0)}(i \omega_n)$ at the first two Matsubara
frequencies.  We find $T_\mathrm{FS}$ from the zero of the effective dispersion
at the antinode $\tilde{\epsilon}_{(\pi, 0)}$. 

Finally, the real-frequency spectral function $A_\bk (\omega)$ and the
self-energy $\Sigma_\bk(\omega)$ are found with the maximum entropy
analysis~\cite{bergeron2016,jarrell1996} on the Green's function
$G_\bk(i\omega_n)$ and the self-energy $\Sigma_\bk(i\omega_n)$.  We use two
independent maximum entropy codes~\cite{bergeron2016,jarrell1996} to make sure
that their results agree.

\section{Supporting material for the results shown in the main text}

\subsection{Pseudogap onset temperature $T^{*}$ and Fermi surface topology transition temperature $T_{FS}$}

\begin{figure}
  \begin{center}
    \includegraphics[scale=0.34]{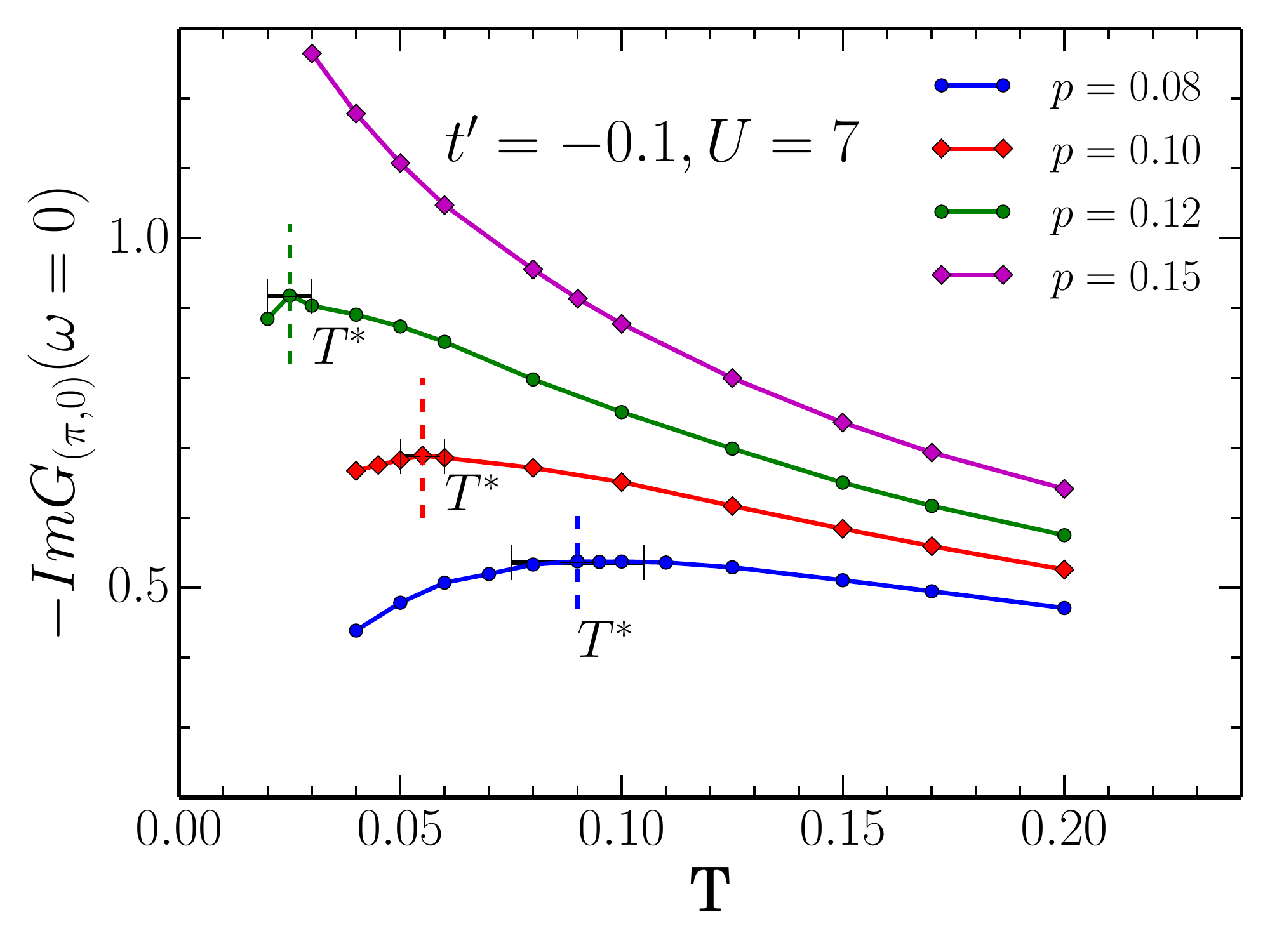}
    \includegraphics[scale=0.34]{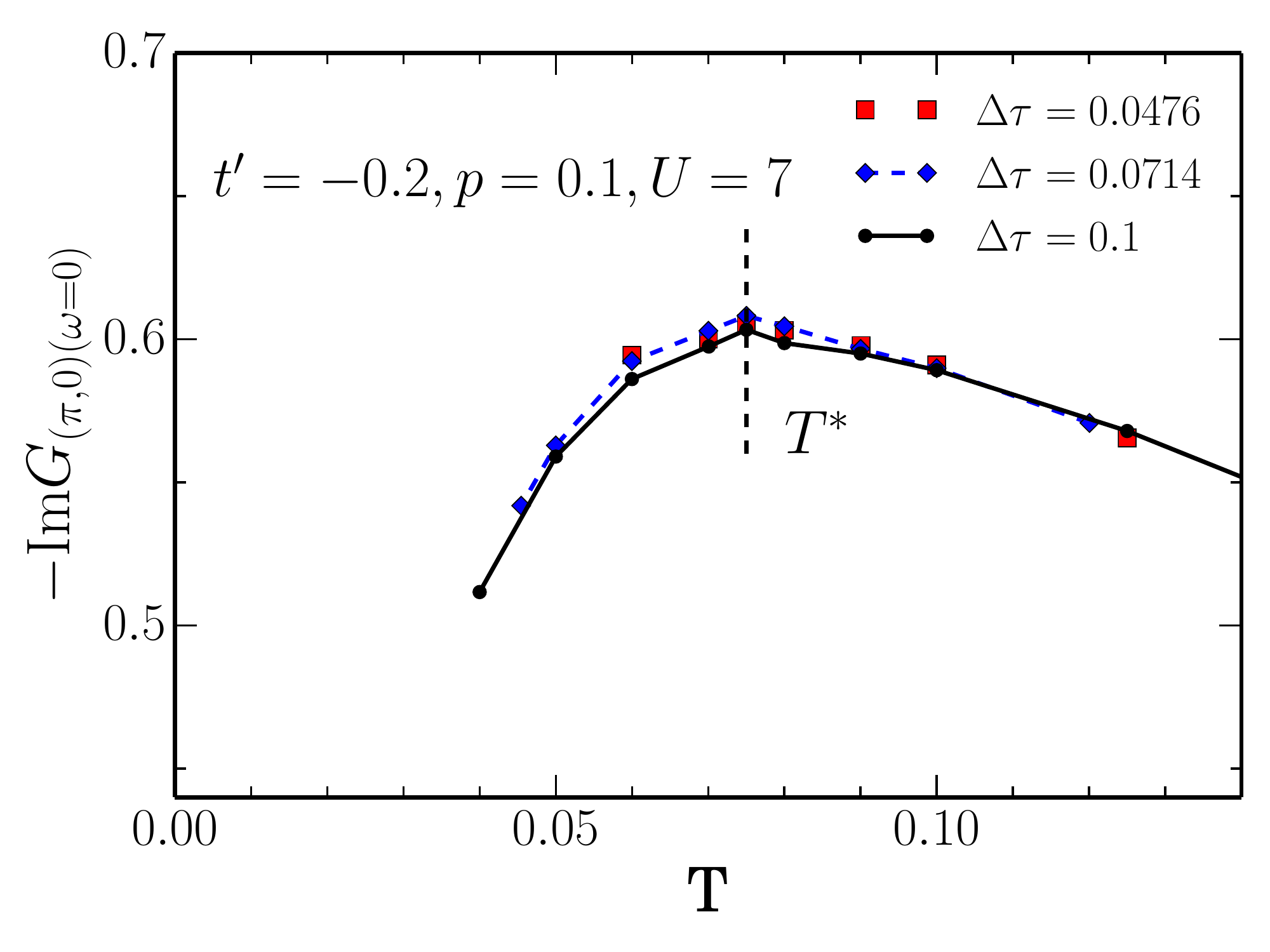}
  \end{center}
  \vspace{-0.4cm}
  \caption{
$-\im G_{(\pi,0)}(\omega=0)$ as a function of temperature $T$. The maximum is identified as the pseudogap onset temperature $T^{*}$. 
\textbf{Upper panel:} for $t^{\prime} = -0.1$ and different doping levels. 
\textbf{Lower panel:} for $t^{\prime} = -0.2, p= 0.1$ and different discrete 
time steps $\Delta \tau$ of the Hirsch-Fye impurity solver. 
We can see that $T^{*}$ is already converged 
for $\Delta \tau = 0.1$.
  \label{Tstar}}
\end{figure}

We identify the pseudogap onset temperature $T^{*}$ as the temperature where
the spectral function at the antinodal point $A_{(\pi,0)}(\omega = 0)$ reaches
a maximum as temperature is lowered, see Fig.~\ref{Tstar}. 
The zero-frequency value of the spectral function
$A_{(\pi,0)}(\omega = 0) = -\frac{1}{\pi} \im G_{(\pi,0)}(\omega=0)$ which is
obtained by a linear extrapolation of the value of $\im G_{(\pi,0)}(i\omega_n)$
(the result of the numerical calculation) at its first two Matsubara
frequencies. We found that different approximations of the spectral function,
such as $A_{(\pi,0)}(\omega=0) \simeq \beta G_{(\pi,0)}(\tau =
\frac{\beta}{2})$~\cite{trivedi1995} yield the same values of $T^*$.
Also, we have used different imaginary time discretization steps $\Delta \tau$
in the Hirsch-Fye algorithm and observe that the results are the same for all
values of $\Delta \tau \le 0.1$.  We have used $\Delta \tau = 0.0476$ throughout
our work.

\begin{figure}
  \begin{center}
    \includegraphics[scale=0.36]{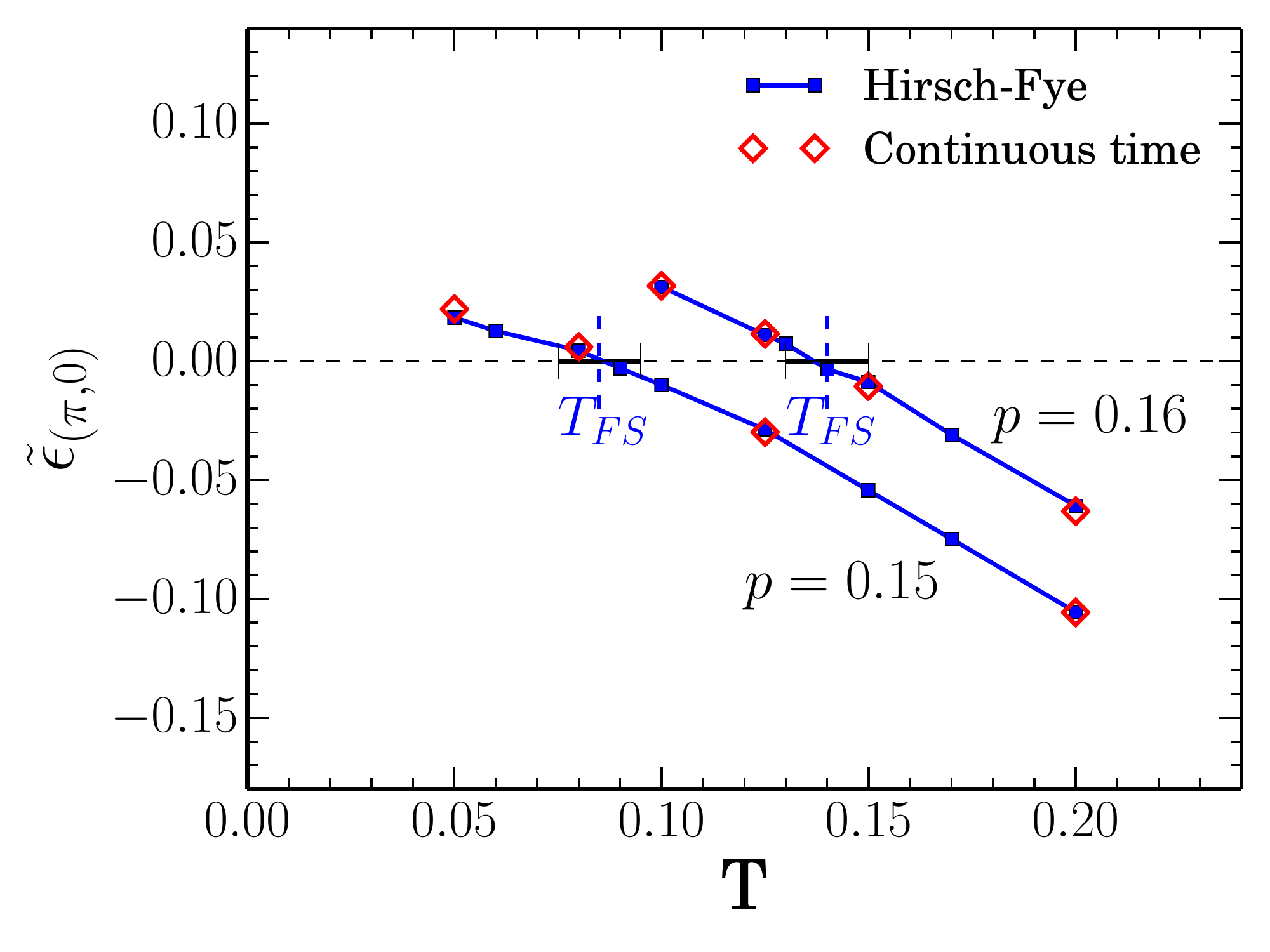}
  \end{center}
  \vspace{-0.4cm}
  \caption{
The effective quasiparticle dispersion at the antinode
$\tilde\epsilon_{(\pi,0)}$ as a function of temperature $T$ for two
doping levels. At the Fermi surface topology transition temperature $T_{FS}$,
we have  $\tilde\epsilon_{(\pi,0)} = 0$.  We show results using both a
continuous-time QMC~\cite{rubtsov2005} 
and a Hirsch-Fye~\cite{hirschfye} impurity solver 
(the latter with a finite imaginary-time step $\Delta \tau  = 0.0476$). 
  \label{Tfs}}
\end{figure}

\begin{figure*}
  \begin{center}
    \includegraphics[scale=0.25]{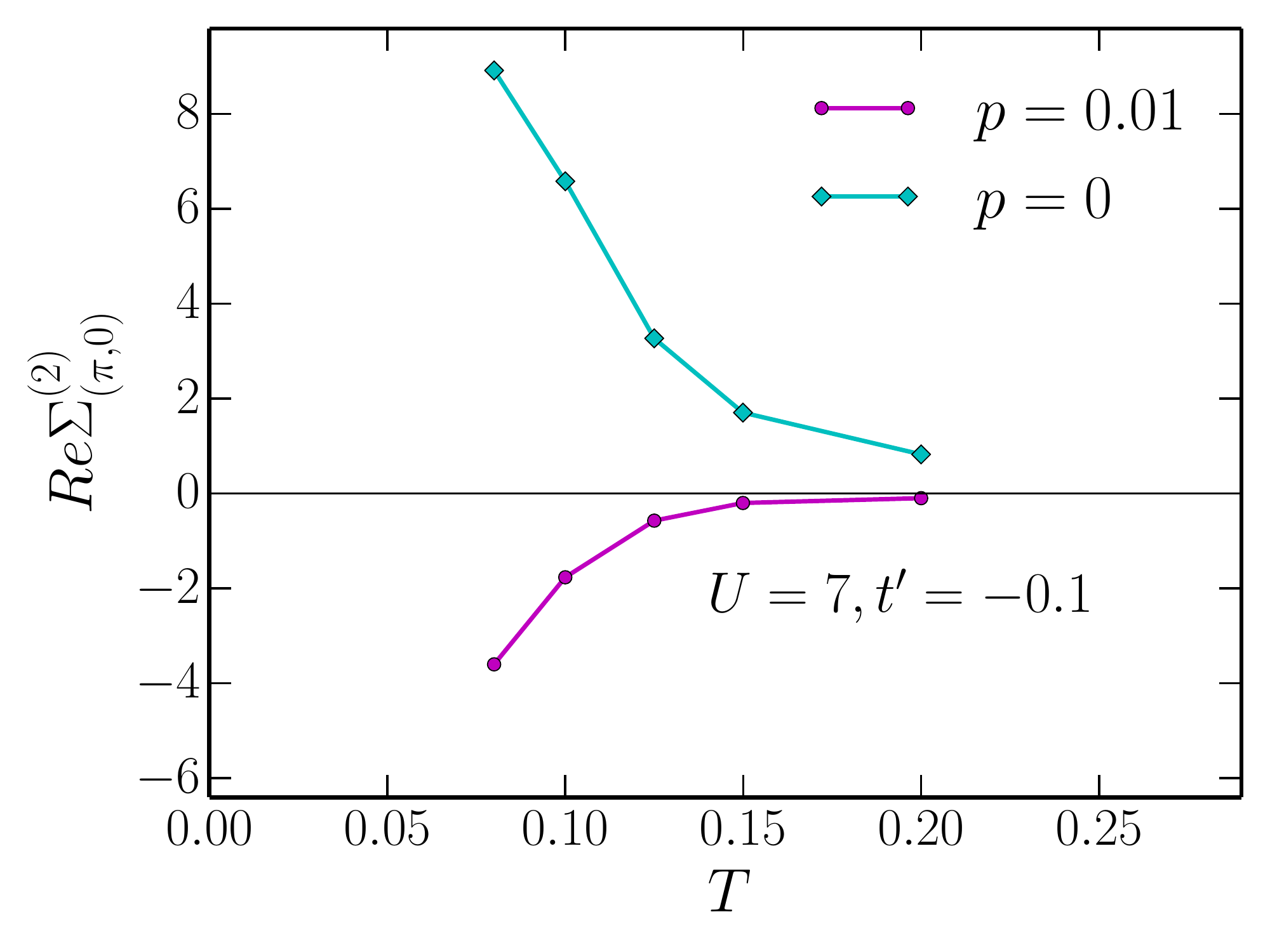}
    \includegraphics[scale=0.25]{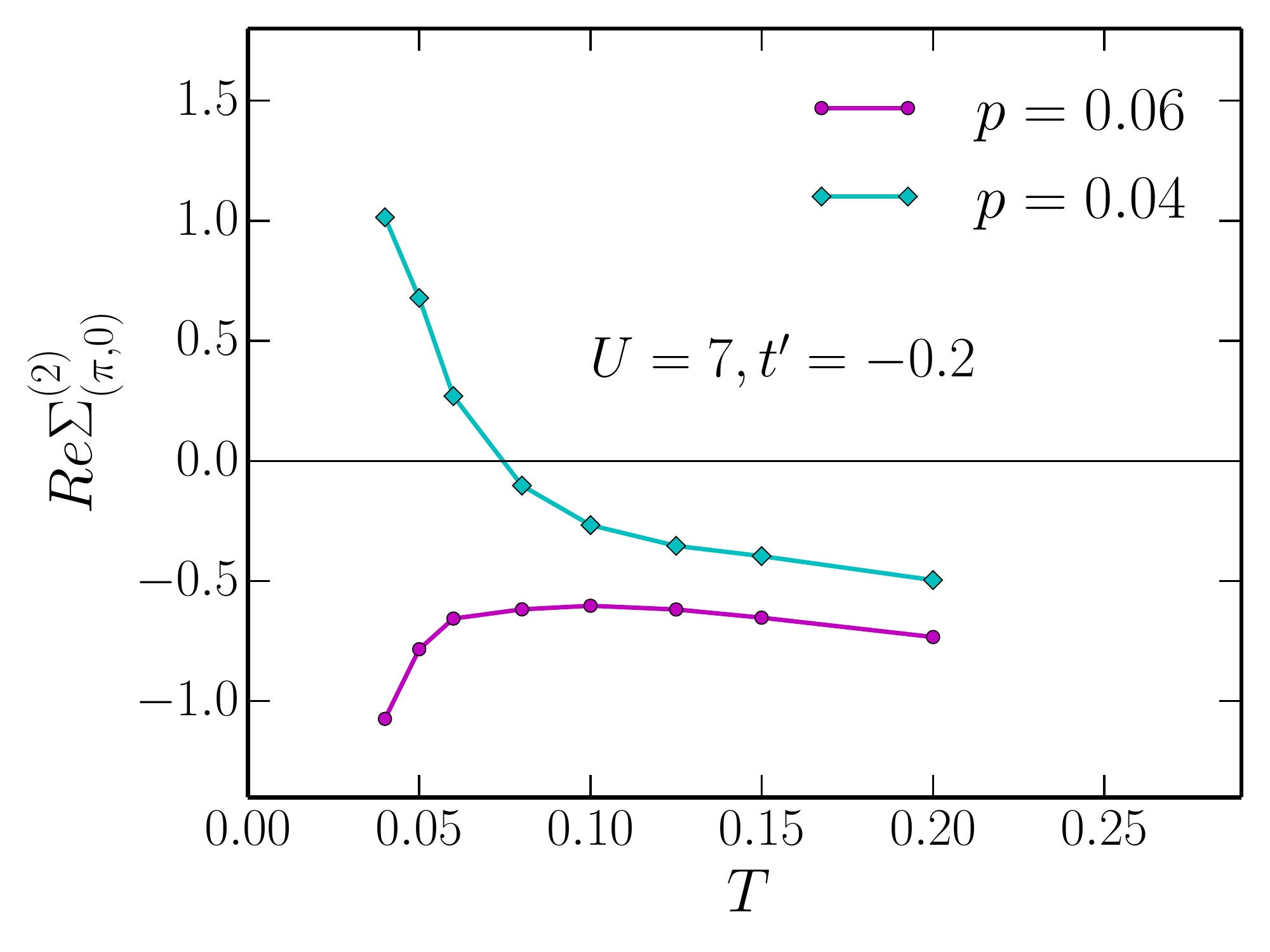}
    \includegraphics[scale=0.25]{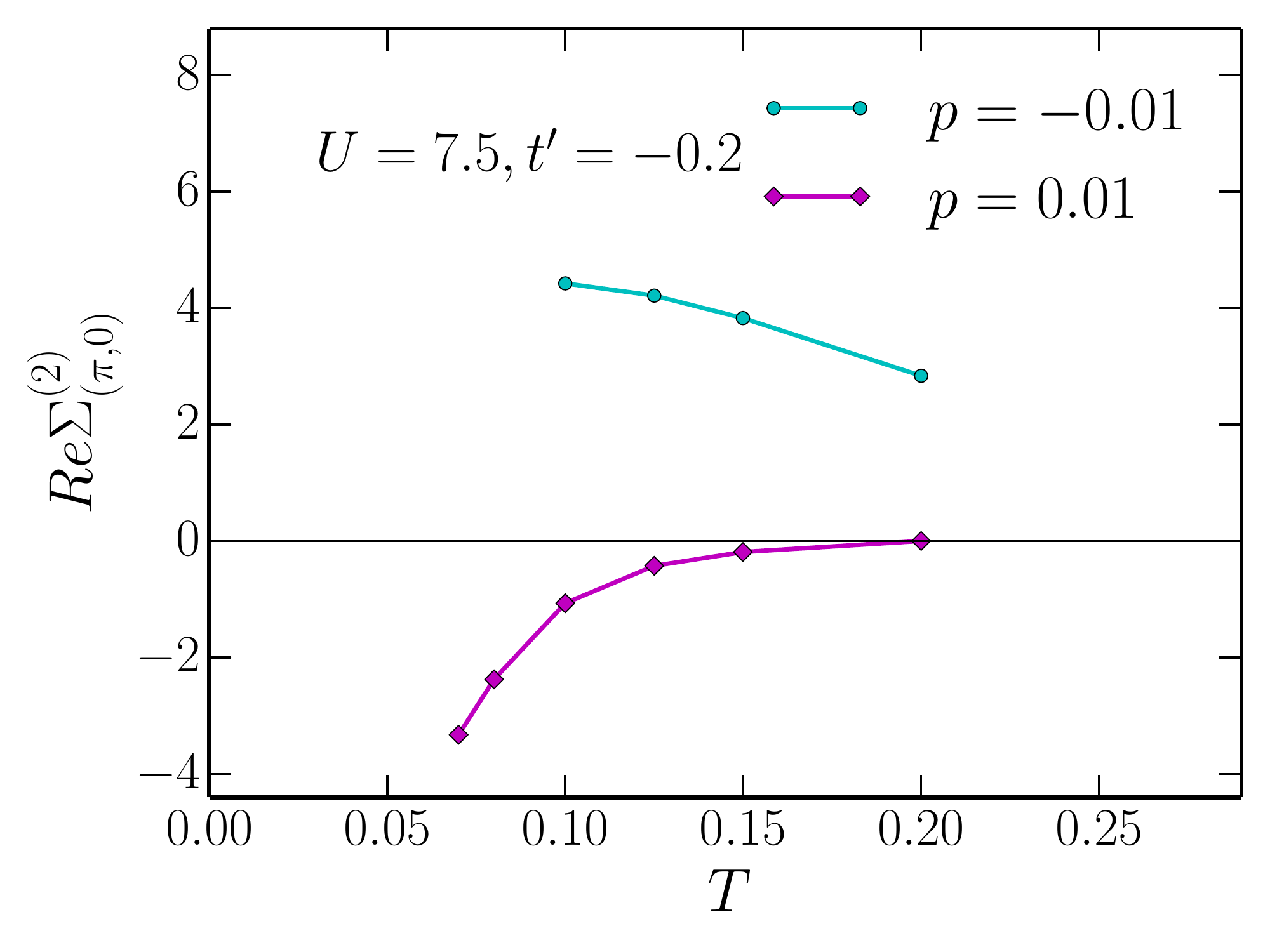}
  \end{center}
  \vspace{-0.4cm}
  \caption{
The real part of the antinodal self-energy as a function of temperature for
different dopings $p$ and $t^{\prime}$. Note that the constant Hartree shift
has been removed. \textbf{Left panel}:  For $U=7, t^{\prime} = -0.1$, at $p=0$, $\re
\Sigma_{(\pi,0)}^{(2)}(\omega=0)$ is positive at low temperatures while at
$p=0.01$ it is negative, suggesting $p_c$ is between 0 and 0.01.  \textbf{Central
panel}: For $U=7, t^{\prime} = -0.2$, we have $0.04<p_c<0.06$. 
\textbf{Right panel}: For $U=7.5, t^{\prime} = -0.2$, we find $-0.01<p_c<0.01$. Therefore for fixed $t^{\prime}$, increasing $U$ drives
$p_c$ closer to the 0 (half-filling).
  \label{redcurve}}
\end{figure*}

When the Fermi surface crosses $k = (\pi,0)$ it undergoes a Lifshitz transition
and changes from hole-like to electron-like. We define the Fermi
surface at finite temperature by the location of the maximum spectral intensity
at zero energy (as seen in an ARPES experiment). This maximum goes through
$(\pi,0)$ when the quasiparticle effective dispersion at the antinode
$\tilde\epsilon_{(\pi,0)} = 0$ as discussed in the main text (see Equation~1).
We can therefore find the Fermi surface topology transition temperature
$T_{FS}$ by finding the zero of $\tilde\epsilon_{(\pi,0)}$ as a function of
temperature as shown in Fig.~\ref{Tfs}. We also display the results as
obtained using two different impurity solvers and show that they give identical
results.

\subsection{Maximum low-energy scattering line and particle-hole asymmetry}

As we have shown in the main text, there is a curve in the $p - t^{\prime}$
diagram that separates a region where the pole-like feature in the imaginary
part of the antinodal self-energy is on the negative energy side from a region where it
is on the positive energy side. When the pole is on the positive energy side,
the real part of the self-energy (with the Hartree term removed) $\re
\Sigma^{(2)}_{(\pi,0)}(\omega=0)$ is negative, while it is positive when the pole is on
the negative side. We can therefore locate the curve by finding, at fixed $t'$,
the value of the doping $p_c$ at which the zero-temperature extrapolation of $\re
\Sigma^{(2)}_{(\pi,0)}(\omega=0)$ changes sign see Fig.~\ref{redcurve}.

\subsection{$U$ dependence of the connection between $p^*$ and $p_{FS}$}

\begin{figure}
  \begin{center}
    \includegraphics[scale=0.4]{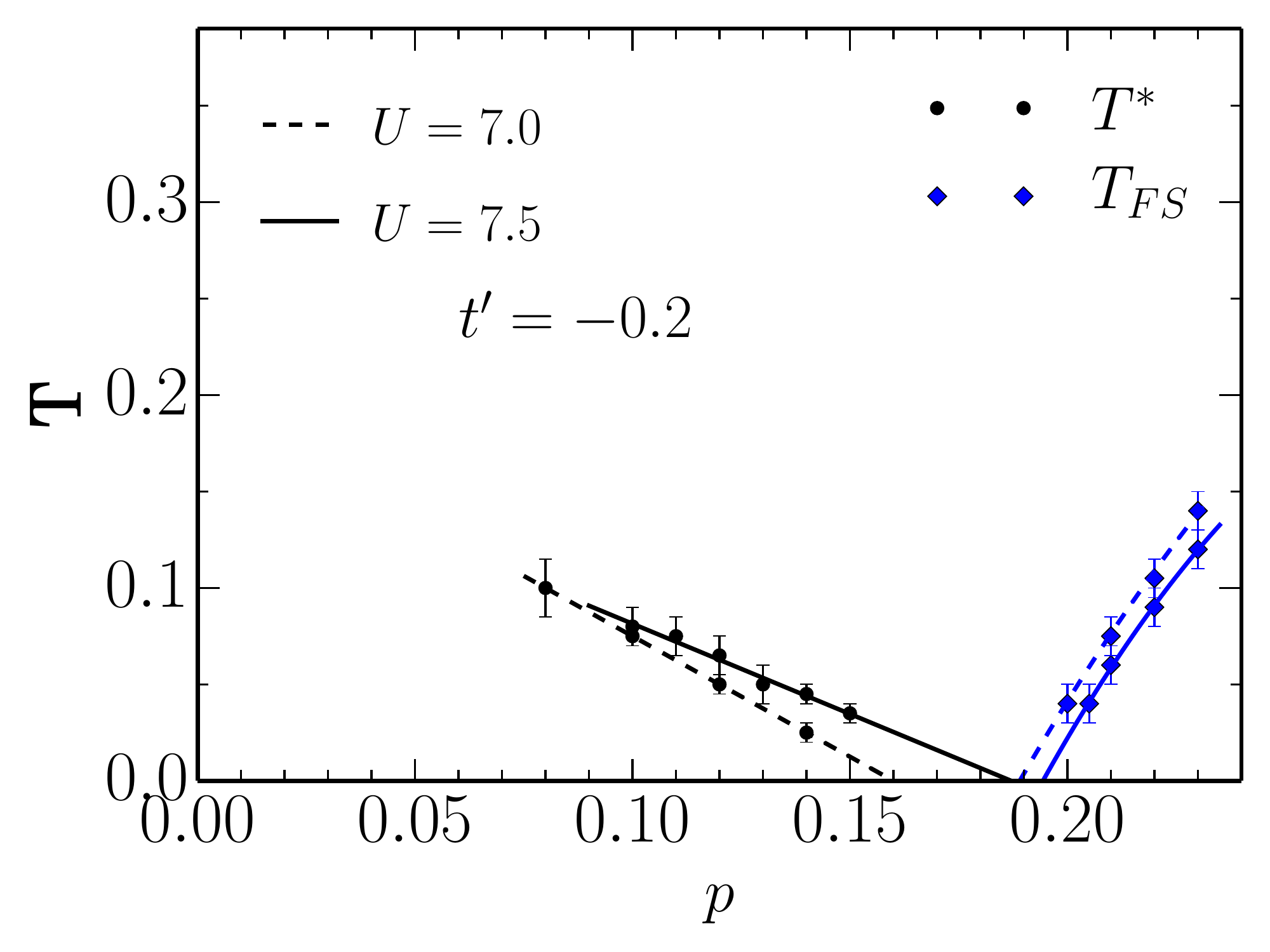}
  \end{center}
  \vspace{-0.4cm}
  \caption{ $T^{*}$ and $T_{FS}$ as a function of doping $p$ at $t^{\prime} = -0.2$, $U=7$ (dashed lines), and $U=7.5$ (solid lines).
  Note that here the imaginary-time discretization step is $\Delta \tau = 1/2U$.
  \label{u7u75}}
\end{figure}

In the main text, all calculations have used $U=7$. For this value, we have
shown that, $p^* \simeq p_\mathrm{FS}$ for values of $t'$ greater than $\approx
-0.1$. For more negative values of $t'$ the $p^*$ and $p_\mathrm{FS}$ lines
split apart. When $U$ is larger, this branching point goes to lower values of
$t'$. This is shown in Fig.~\ref{u7u75} where we compute $T^*$ and
$T_\mathrm{FS}$ for both $U=7$ and $U=7.5$. It is clear from the figure, that
for $U=7.5$ $p^*$ and $p_\mathrm{FS}$ are much closer than for $U=7$. This can
be understood because a larger value of $U$ extends the pseudogap region to
larger dopings, while the Fermi surface topology is not influenced much by
correlations before we actually have a pseudogap and hence $p^* \simeq
p_\mathrm{FS}$.

\section{Temperature evolution of $\tilde\epsilon_k$ and role of non-local
correlations}

\begin{figure}
  \begin{center}
    \includegraphics[scale=0.4]{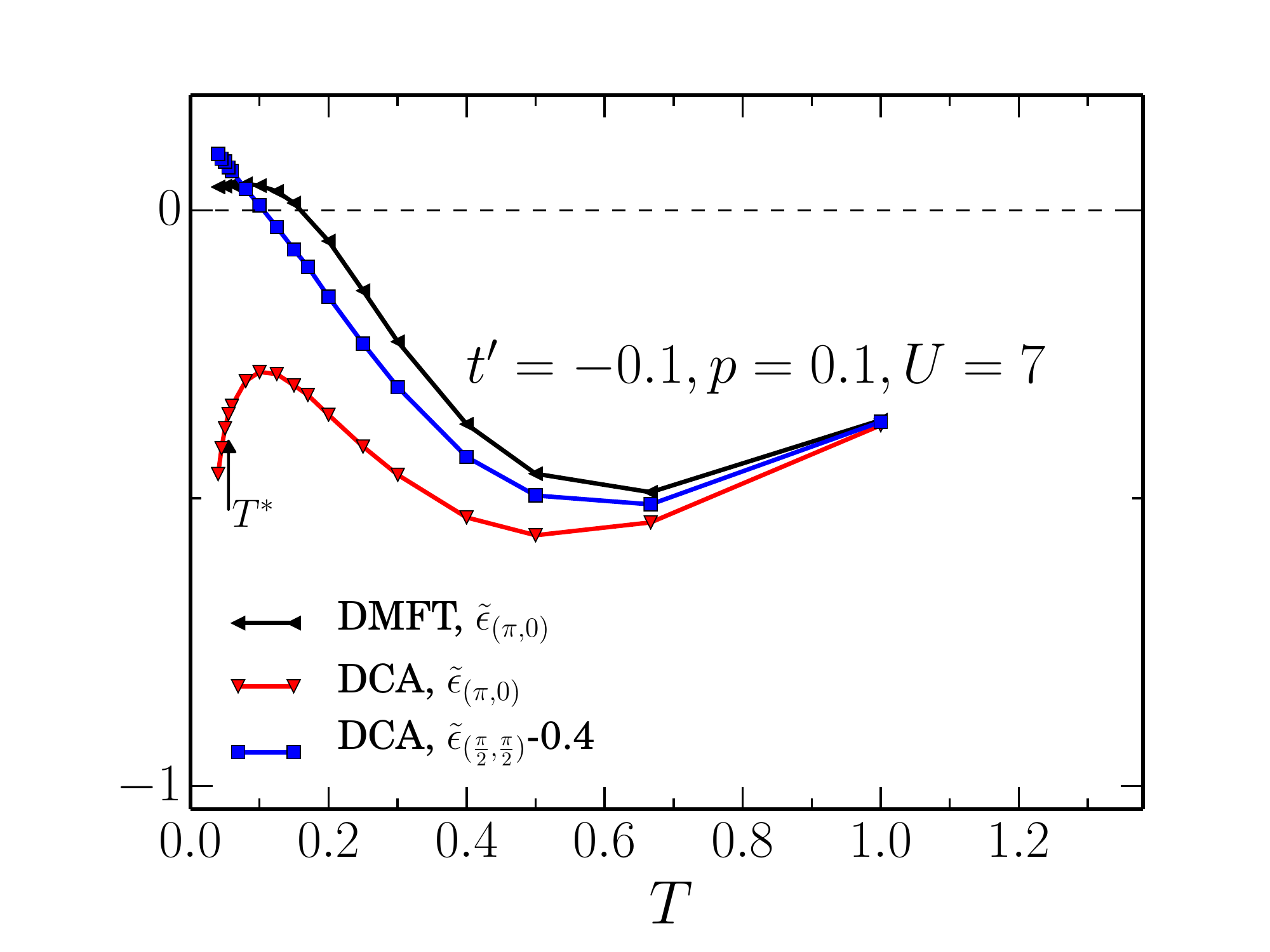}
  \end{center}
  \vspace{-0.4cm}
  \caption{
Temperature dependence of $\tilde\epsilon_k$ as obtained by DCA and DMFT at
$10\%$ hole doping and $t' = -0.1$.
  \label{ekdmft}}
\end{figure}

In Fig.~\ref{ekdmft}, we investigate the role of non-local correlations by
comparing the results obtained by DCA, as in the main text, and single-site
dynamical mean-field (DMFT) that only accounts for local correlations.  The
black line show $\tilde\epsilon_{(\pi,0)}$ as computed by DMFT for $t'=-0.1$
and $p=0.1$. For these parameters, the non-interacting Fermi surface is
electron-like. It is seen that at low temperature the DMFT results also predict
an electron-like Fermi surface. This is not surprising as DMFT preserves the
Luttinger theorem and the interacting Fermi surface is the same as the
non-interacting one when $T \rightarrow 0$. However, as temperature is
increased, $\tilde\epsilon_{(\pi,0)}$ decreases significantly and becomes
negative. This yields a hole-like interacting Fermi surface at high temperature
that breaks Luttinger's theorem~\cite{deng2013,shastry2011} with a volume
larger than in the non-interacting case.

The red and blue lines show $\tilde\epsilon_{(\pi,0)}$ and
$\tilde\epsilon_{(\frac{\pi}{2},\frac{\pi}{2})}$ respectively as obtained by
DCA. $\tilde\epsilon_{(\frac{\pi}{2},\frac{\pi}{2})}$ has been shifted by a
constant $4t' = -0.4$ that corresponds to the energy difference of the
non-interacting dispersion at $(\pi,0)$ and $(\frac{\pi}{2},\frac{\pi}{2})$. At
high temperatures all curves yield the same value, compatible with a
self-energy that is essentially local. As temperature is decreased the nodal
$\tilde\epsilon_{(\frac{\pi}{2},\frac{\pi}{2})}$ behaves like the DMFT solution
indicating that the Fermi surface at the node is very close to its
non-interacting shape. The DCA $\tilde\epsilon_{(\pi,0)}$ has a different
behavior. As temperature is lowered it quickly departs from
$\tilde\epsilon_{(\frac{\pi}{2},\frac{\pi}{2})}$ showing the onset of
nodal/antinodal differentiation. At a temperature slightly above $T^*$,
non-local correlations become large and induce a very negative
$\tilde\epsilon_{(\pi,0)}$ as discussed in the main text.

\section{Pseudogap and Fermi surface topology transition as a function of $U$}

\begin{figure}[!ht]
  \vspace{0.4cm}
  \begin{center}
    \includegraphics[scale=0.4]{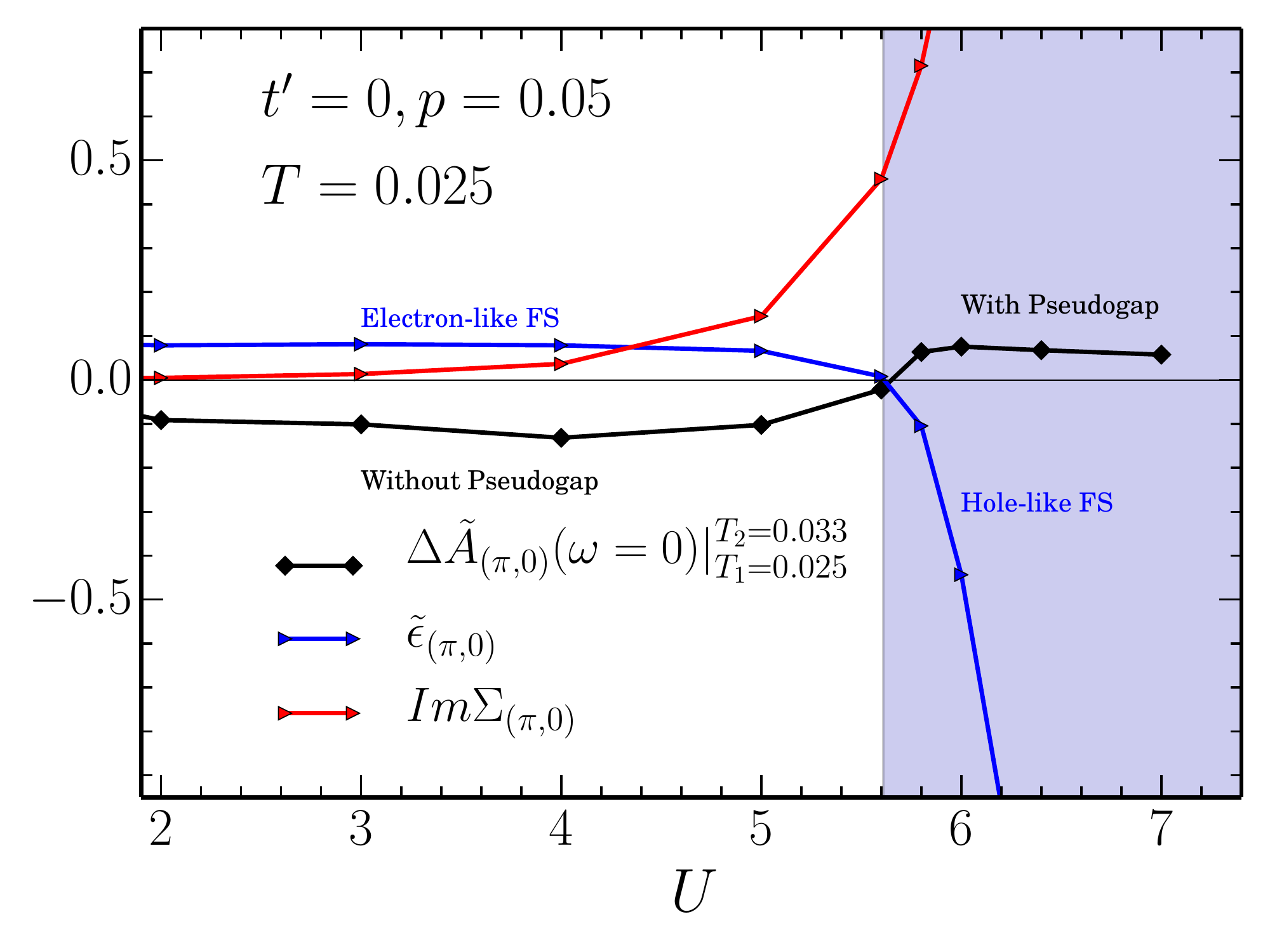}
  \end{center}
  \vspace{-0.4cm}
  \caption{
Correlation effects shown as a function of $U$ for three quantities: the
quasiparticle effective dispersion $\tilde\epsilon_{(\pi,0)}$, the imaginary
part of the antinodal self-energy at zero energy $\im
\Sigma_{(\pi,0)}(\omega=0)$ and the difference in spectral intensity at the
Fermi level for the two lowest calculated temperatures indicating whether a
pseudogap  has formed. 
  \label{vsU}}
\end{figure}

Our results for $U=7$ show that, for a broad range of parameters, the
pseudogap disappears at the same critical doping where the Fermi surface
undergoes a Lifshitz transition. In Fig.~\ref{vsU}, we investigate how
our results depend on the correlation strength $U$. It is shown that for
values of $U \lesssim 5$ correlations have little effect on the Fermi
surface topology and the low-energy scattering rate $\im \Sigma_{(\pi,0)}$
is very small. Above $U \simeq 5$ correlation effects set in quickly as shown
by a fast increase in the value of $\im \Sigma_{(\pi,0)}$. This induces
a pseudogap at $U = 5.6$. At the same time
the effective quasiparticle dispersion $\tilde\epsilon_{(\pi,0)}$ crosses
zero and becomes very negative for larger values of $U$. This sudden
increase of the correlation effects for $U > 5$ might explain why the
pseudogap and the Fermi surface topology happen at the same time.

\section{Comparison with weak-coupling approaches}
\label{app:weak}

Let us investigate how our results differ from weak-coupling approaches, such as spin-fluctuation theory or 
the two-particle self-consistent approach (TPSC) of Vilk and Tremblay~\cite{vilk1997}. 
The latter has been shown to be quite accurate in the weak to intermediate coupling regime 
of the two-dimensional Hubbard model, and this appendix closely follows the analysis
in Ref.~\onlinecite{vilk1997}. 

In those approaches, the self-energy is obtained as
\begin{equation}
  \Sigma(\bk, i \omega_n) = g^2\, T\sum_p \frac{1}{V_{\mathrm{BZ}}}\sum_\bq  G_0(\bk+\bq, i \omega_n + i \nu_p) \chi(\bq, i \nu_p),
\end{equation}
where $G_0$ is the non-interacting Green's function, $\chi$ is the spin susceptibility, and 
$g$ is a coupling constant with the dimension of energy.
When the magnetic correlation length $\xi$ is large, $\chi$ can be approximated by:~\cite{tremblay2012}
\begin{equation}
  \chi(\bq, i\nu_p) \propto \frac{1}{(\bq-\bQ)^2 + \xi^{-2} \nu_p / \omega_\mathrm{sp} + \xi^{-2}},
\end{equation}
with $\bQ = (\pi,\pi)$ the antiferromagnetic wave-vector. 
The self-energy thus reads
\begin{align}
\begin{split}
 \Sigma(\bk, i \omega_n) \propto T\sum_p \frac{1}{V_{\mathrm{BZ}}}\sum_\bq  
 \frac{1} {i\omega_n + i\nu_p + \mu - \epsilon_{\bk+\bq}} \\
  \times  \frac{1}{(\bq-\bQ)^2 + \xi^{-2} \nu_p / \omega_\mathrm{sp} + \xi^{-2}}.
\end{split}
\end{align}
In the regime of interest here (large enough $\xi$, renormalized classical regime), the above sum is dominated 
by the smallest Matsubara frequency (note that $\omega_\mathrm{sp} \sim \xi^{-2}$) and one obtains 
the imaginary part of the retarded real-frequency self-energy in the form:   
\begin{equation}
-\frac{1}{\pi} \mathrm{Im}\, \Sigma_{ret}(\bk,\omega) 
\propto T\,\int d^2\bq\, \delta(\omega-\xi_{\bk+\bq})\,\frac{1}{(\bq-\bQ)^2+\xi^{-2}}
\end{equation}
The important point is that in two dimensions, this integral diverges as the correlation length becomes 
large, which leads to the formation of `hot spots' at which a pseudogap opens. This integral can actually 
be performed analytically, and one finally obtains:
\begin{equation}
-\frac{1}{\pi} \mathrm{Im}\, \Sigma_{ret}(\bk,\omega)\,=\,
\tilde{g}\, \frac{T}{\sqrt{(\omega-\xi_{\bk+\bQ})^2+(v_F/\xi)^2}} + \mathrm{reg.}
\label{eq:self_sf}
\end{equation}
where `reg.' denotes a non-singular contribution.  
The physics associated with a weak-coupling description of spin fluctuations can be entirely described 
on the basis of this expression ~\cite{vilk1997}. Let us focus first on the Fermi surface properties, corresponding to $\omega=0$ and 
momenta such that $\xi_\bk=0$. As clear from (\ref{eq:self_sf}), the self-energy is regular on the Fermi surface except at the `hot spots' 
satisfying also $\xi_{\bk+\bQ}=0$, corresponding to the intersection of the Fermi surface with the antiferromagnetic 
Brillouin zone. At these hot-spots, the self-energy is singular: its imaginary part is of order:
\begin{equation}
-\frac{1}{\pi} \mathrm{Im}\,\Sigma|_{hot} \propto \frac{T\xi}{v_F} 
\end{equation}
This is large only when the correlation length is large: $\xi > v_F/T$. In this regime, spectral weight is strongly 
depleted at the hot spots, corresponding to the weak-coupling description of the pseudogap. When the correlation 
length remains finite as temperature is lowered (short-range order), the hot spots and corresponding pseudogap disappear 
for $T<v_F/\xi$ and conventional Fermi liquid behavior is recovered at low temperature. 

Let us emphasize the crucial differences that exist between the weak-coupling expression of the self-energy (\ref{eq:self_sf}) 
and both the self-energy that we obtain from DCA at strong coupling, as well as the self-energy obtained from the 
SU(2) gauge theory. As clear from (\ref{eq:self_sf}), the imaginary part of the weak-coupling self-energy does display a peak, 
but (i) the height of this peak is proportional to $T\xi(T)$ and thus eventually the peak and the hot spots disappear at low-T if 
$\xi$ remains finite (ii) the width of this peak is proportional to $v_F/\xi < T$, 
which in the regime where the peak exists is smaller than temperature. In contrast, in the strong-coupling DCA calculations 
the peak is not suppressed as $T$ is reduced, and its width is larger than $T$. 
Furthermore, the correlation length that we can estimate in our DCA results from the static staggered susceptibility $\chi_{AF}\propto \xi^2$ is quite 
small at strong coupling: we find for example: $\xi/a\simeq 2.7$ for $U=7, t^\prime=-0.2, p=0.1$ at $T=1/30$. 
The weak-coupling expression also has a different structure than the singular delta-function form of the chargon 
self-energy in the SU(2) gauge theory: the latter, importantly, does not involve the correlation length (set by the spinons) 
and is similar to that of an SDW in the ordered phase. 

\begin{figure}
  \begin{center}
    \includegraphics[scale=0.4]{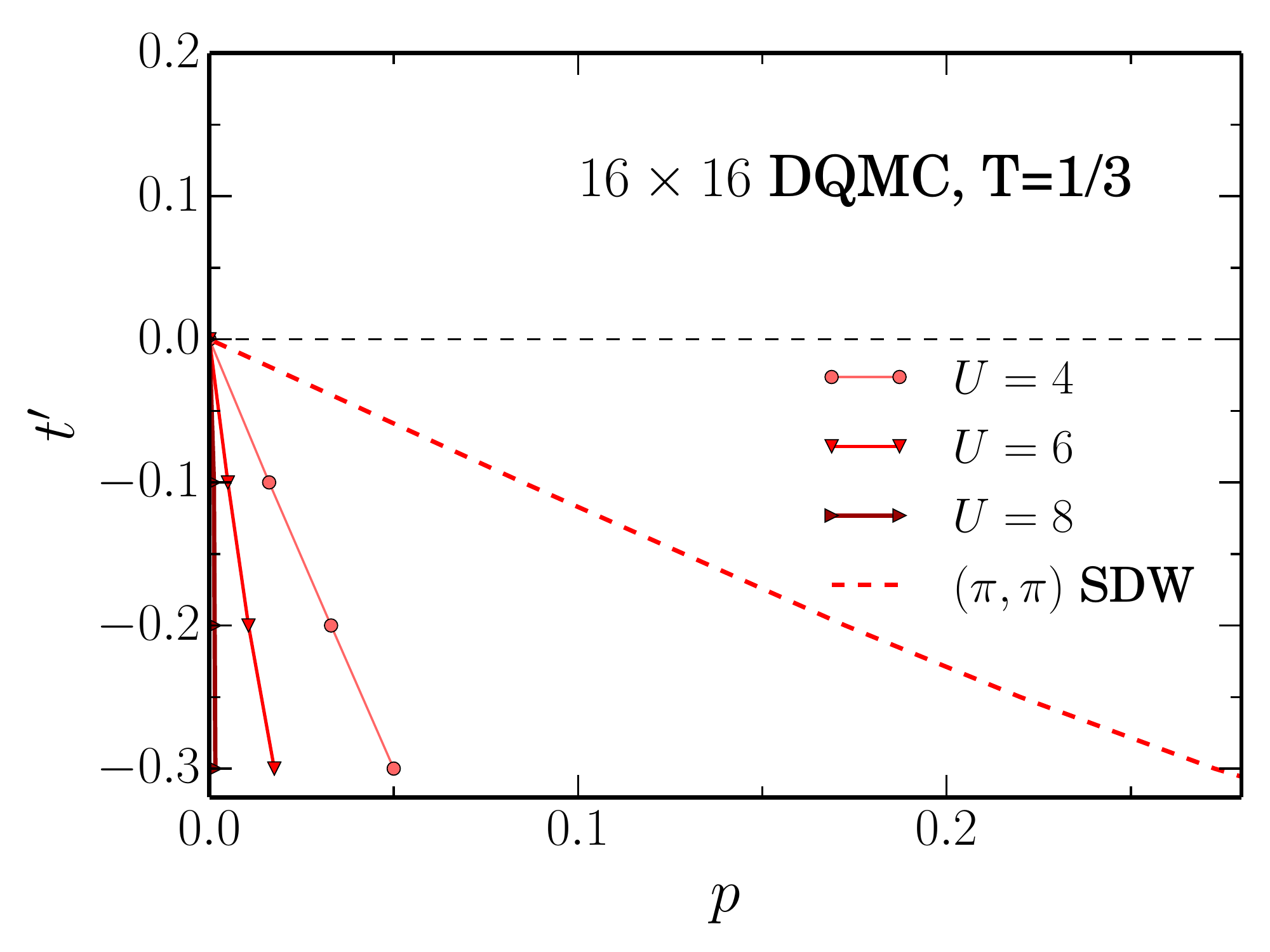}
  \end{center}
  \vspace{-0.4cm}
  \caption{
The lines in the $t'-p$ plane show where the low-energy imaginary part of the
antinodal self-energy $\im \Sigma_{(\pi,0)}(\omega)$ has its pole-like feature
centered around $\omega=0$ and an essentially particle-hole symmetric
low-energy spectrum. On these curves $\re \Sigma_{(\pi,0)}(\omega=0)$ vanishes.
The solid lines are obtained by DQMC for different values of $U$, while the
dashed line is the result from the SDW weak-coupling approach. The dashed line
coincides with the non-interacting Lifshitz transition.
  \label{symmetry2}}
\end{figure}

The real part of the self-energy corresponding to (\ref{eq:self_sf}) can be obtained using Kramers-Kronig 
relations as:
\begin{align}
\begin{split}
\mathrm{Re}\, \Sigma_{ret}(\bk,\omega)\,\propto \, &
\frac{T}{\sqrt{\Omega_k^2+(v_F/\xi)^2}}\,
\\
&\times
\ln |\frac{\Omega_k+\sqrt{\Omega_k^2+(v_F/\xi)^2}}
{\Omega_k-\sqrt{\Omega_k^2+(v_F/\xi)^2}}|
\end{split}
\end{align}
in which we have used the short-hand notation $\Omega_k\equiv \omega-\xi_{\bk+\bQ}$. 
In the temperature regime where hot spots are present $T>v_F/\xi$, one can distinguish two regimes 
of frequencies. 
For $\Omega_k > v_F/\xi$, $\mathrm{Re}\Sigma$ is of order $(T/\Omega_k) \ln(\Omega_k\xi/v_F)$, while 
at low frequencies $\Omega_k < v_F/\xi$, the self-energy is regular 
$\mathrm{Re}\Sigma\propto T\Omega_k (\xi/v_F)^2$. 
Hence, the regular part of the self-energy is regular at low-frequency even close to the hot spots when $\xi$ remains finite.  
As a result, hot spots exist at the intersection between the Fermi surface and the antiferromagnetic Brillouin zone for 
$T>v_F/\xi$, but there is no reconstruction of the Fermi surface otherwise. 
Hence, for a doping value larger than the value corresponding to the non-interacting Lifshitz transition, there are 
no hot spots and the Fermi surface is weakly renormalized and electron-like. 
Hence, in weak-coupling the non-interacting Lifshitz transition controls both the 
location of the self-energy singularities and the topological transition of the Fermi surface. 
This is very different from our results in the strong coupling regime $U=7$ where these phenomena are
controlled by three different lines. 

By varying $U$ one can observe how the
transition from weak to strong coupling happens.  Fig.~\ref{symmetry2} shows
the DQMC results for several values of $U$. The lines show where the real part
of the self-energy vanishes. It separates a region where the pole in the
self-energy is at negative energies and one where it is on the positive side.
It is seen that as $U$ becomes smaller the lines slowly approach the
non-interacting Lifshitz transition, as expected in weak-coupling.

\section{Comparison of the chargon and electron self-energy in the SU(2) gauge theory}
\label{app:convolution}

\begin{figure}
  \begin{center}
    \includegraphics[scale=0.45]{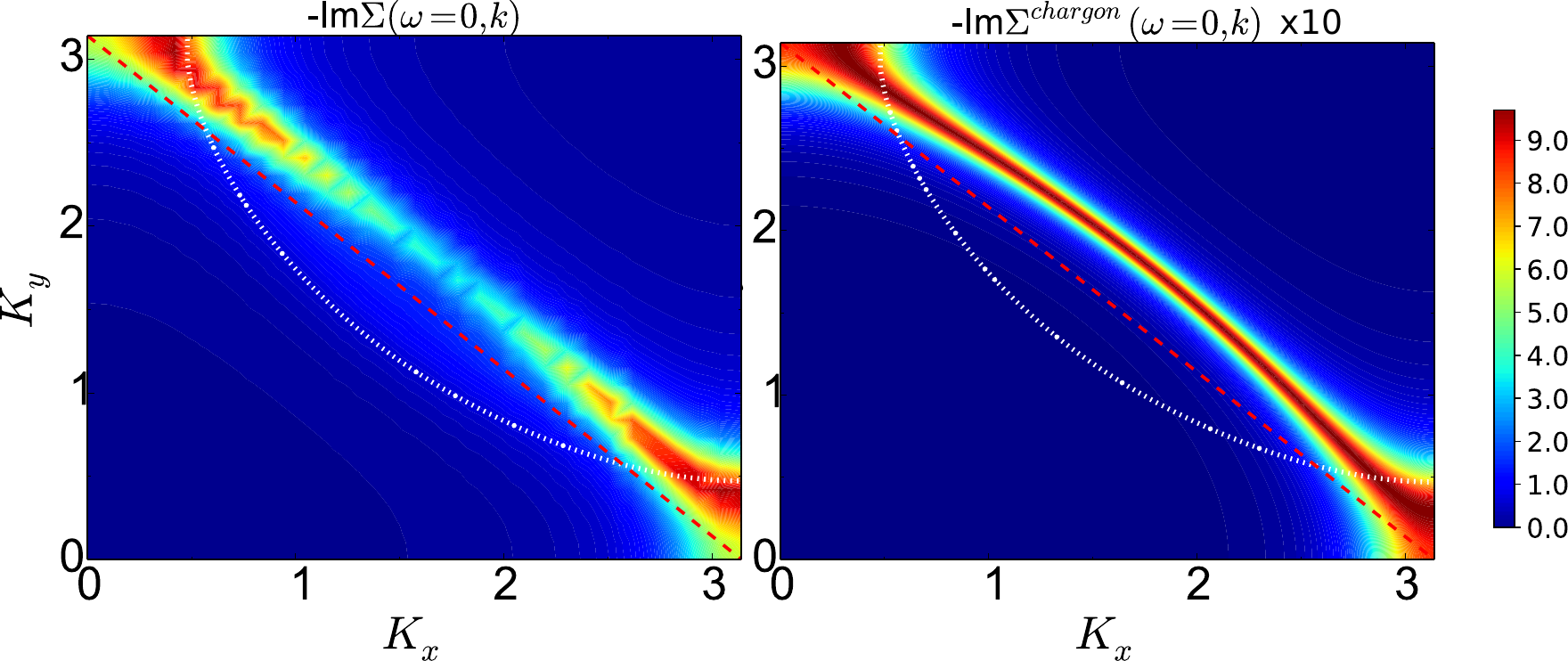}
  \end{center}
  \vspace{-0.4cm}
  \caption{
 Scattering rate
of the electrons $- \im \Sigma_{k}(\omega=0)$ and chargons $-\im
\Sigma^{chargon}_{k}(\omega=0)$ in momentum space. The parameters used here are: $H_0 = 0.2, J = 0.1; T = 1/30,  t' = -0.3,  Z_{t} = 0.31, Z_{t'}=0.19, \Delta = 0.01, p = 0.05$. \textbf{Left panel:} The
physical electron self-energy has clear nodal/antinodal differentiation with
a stronger scattering at the antinode than at the node. \textbf{Right panel:}
The chargon self-energy is given by Eq.~\ref{eq:GreensFunctionChargon} and
has no momentum differentiation. 
The broadening in the chargon self-energy is $\eta = 0.04$.
  \label{fig:su2sigma}}
\end{figure}

Here we illustrate in more details the role of the convolution that allows to
recover the electronic Green's function in the SU(2) theory. As we have
discussed above, see e.g. Fig.~\ref{fig:gauge}, the location in momentum and
frequency of the most singular structures of the physical self-energy are not
affected by the convolution and they are already encoded in the chargon
self-energy given by Eq.~\ref{eq:GreensFunctionChargon}. The convolution mainly
smears $G_\psi$ and the electron self-energy is a broadened counterpart of the
chargon self-energy. A more detailed inspection shows that the convolution also
redistributes spectral weight over the Brillouin zone. As a result, the
physical electron self-energy displays nodal/antinodal differentiation, which
is absent in the chargon self-energy.  This is illustrated in
Fig.~\ref{fig:su2sigma} where it is clearly seen that the imaginary part of the
electronic self-energy is larger close to the antinode than at the node. This
differentiation is not present in the chargon self-energy.

\bibliography{pseudogap}

\end{document}